%% file: Chapter_AKR.tex
\documentclass[graybox,envcountchap,sectrefs,vecphys]{svmono}
\usepackage{type1cm}         

\usepackage{makeidx}         
\usepackage{graphicx}        
                             
\usepackage{multicol}        
\usepackage[bottom]{footmisc}

\usepackage{physics}
\usepackage{amsmath}

\usepackage{amsfonts}

\usepackage{mathrsfs}

\usepackage{bm}

\usepackage{tabto}

\usepackage{mathtools}

\usepackage{amssymb}

\newcommand{\bcdot}{\boldsymbol{\cdot}}

\usepackage{braket}

\usepackage{epigraph}

\makeindex

\usepackage{newtxtext} 
\usepackage[varvw]{newtxmath}       

\begin{document}


\title{Mathematical Foundation for Quantum Computing of 
Electromagnetic Wave Propagation in Dielectric Media}

\maketitle

\frontmatter

\include{affiliations}

\include{abstract}

\tableofcontents

\include{acronym}

\mainmatter

\include{intro}

\include{tensor1}

\include{bra_ket}

\include{quantum_postulates}

\include{bits_qubits}

\include{maxwell}

\include{qla_oned}

\include{conclusions}

\include{appendix_A}

\include{supplementary_mat}

\backmatter

\end{document}

%% file: affiliations.tex
\section*{}


{\Large
\begin{center}\noindent
{\bf Abhay K. Ram} \\
{\it Plasma Science and Fusion Center}\\
{\it Massachusetts Institute of Technology}\\
{\it Cambridge, MA, USA}\\
{abhay@mit.edu}\\
{\it \phantom{this is a test}}\\
{\bf Efstratios Koukoutsis} \\
{\it School of Electrical and Computer Engineering}\\
{\it  National Technical University of Athens}\\
{\it  Zographou, Greece}\\
{stkoukoutsis@mail.ntua.gr} \\
{\it \phantom{this is a test}}\\
{\bf George Vahala} \\
{\it Department of Physics}\\
{\it College of William \& Mary}\\
{\it Williamsburg, VA, USA}\\
{gvahala@gmail.com}\\
{\it \phantom{this is a test}}\\
{\bf Kyriakos Hizanidis}\\
{\it School of Electrical and Computer Engineering}\\
{\it  National Technical University of Athens}\\
{\it  Zographou, Greece}\\
{kyriakos@central.ntua.gr}
\end{center}
}

\vspace{0.5in}
{\Large
\noindent
This chapter is part of the book
\begin{center}
{\bf Emerging Applications of Ions and Plasmas} \\
( Editors:  Samar K. Guharay \& Motoi Wada )\\
\vspace{0.1in}
DOI:\ \ https://doi.org/10.1007/978-3-031-84245-0
\end{center}
}

%% file: abstract.tex
\section*{\LARGE Abstract}

{\Large

Can quantum computers effectively simulate the propagation and scattering of electromagnetic
waves in a classical plasma? This chapter introduces some of the 
basic concepts in mathematics and physics essential to answering that question. 
The numerical simulations of Maxwell equations for wave propagation in dielectrics
are constrained by technological limitations of the present-day computers.
In contrast, there has been ample fanfare around quantum computers and their potential to far exceed the
performance of traditional computers. Whether the enhanced capabilities of a quantum computer can
be put to use for simulating topics in classical physics is a source of intrigue and curiosity.

\vspace{\baselineskip}
\begin{flushright}\noindent
{\it Abhay K. Ram} \\
{\it Efstratios Koukoutsis} \\
{\it George Vahala} \\
{\it Kyriakos Hizanidis}\\
\end{flushright}

}
\vspace{2in}

\noindent
{\bf Keywords}: Quantum computing, quantum information science, linear algebra, vector space and tensors,
Hilbert space, Minkowski space, 
covariant form of Maxwell equations, electromagnetic wave propagation, quantum lattice algorithm.

%% file: acronym.tex
\section*{List of symbols}

\begin{description}[CABR]
\item [$\lambda$]{Characteristic length scale}
\item [$n$]{Particle number density}
\item [$h$]{Planck constant}
\item[$k_B$] {Boltzmann constant}
\item[$T$]{Temperature}
\item [$e$]{Electron charge}
\item[$m$]{Particle mass} 
\item[$\epsilon$] {Electric permeability} 
\item[$\mathcal{F}$]{Field}
\item[$a$] {Field element} 
\item[$\times$] {Cartesian product}
\item[$\in$]{Belongs} 
\item[$\exists$] {Exists}
\item[$\mathbb{R}$]{The field of real numbers}
\item[$\mathbb{C}$]{The field of complex numbers}
\item[$\mathbb{N}$]{The set of natural numbers $\left\{1, 2, 3, \dots \right\}$}
\item[$\subset$]{Proper subset} 
\item[${\bm V}$]{Linear vector space}
\item[$\mathbf v$]{Vector space element}
\item[${\mathbf e}_i$] {Contravariant basis element of vector space} 
\item[${\bm V}^*$]{Linear dual vector space}
\item[$\mathfrak g$]{Dual vector space element}
\item[${\mathbf f}^i$]{Covariant basis element of dual vector space}
\item[$\cap $] {Set-theoretic intersection}
\item[${\mathbf T}$] {Tensor}
\item[$g$]{Metric tensor} 
\item[$\mathscr H$] {Hilbert space}
\item[$\ket{\psi}$] {Ket vector}
\item[$\bra{\psi}$] {Bra vector}
\item[$\mathcal{T}$] {Linear operator}
\item[${\mathbf H}$] {Hamiltonian operator}
\item[$\ket{\Psi}_{B}$] {Bell quantum state}
\item[${\rm CNOT}$] {Controlled NOT operation}
\item[$\mathbf E$] {Electric field}
\item[$\mathbf B$] {Magnetic field}
\item[$\mathbf D$] {Displacement electric field}
\item[$\mathbf H$] {Magnetic intensity}
\item[$\rho$] {Electric charge density}
\item[$\mathbf j$] {Current density}
\item[$\mu$]{Magnetic permeability} 
\item[$\mathbf P$] {Polarization density}
\item[$\mathbf M$] {Magnetization density}
\item[${\mathscr M}$] {Minkowski real vector space}
\item[$\eta$] {Minkowski metric}
\item[$\Lambda$] {Lorentz transformation}
\item[$F^{\mu\nu}$]{Covariant electromagnetic field tensor}
\item[$c$]{Speed of light in the vacuum} 
\item [RSW] {Riemann-Silberstein-Weber}
\item[$\mathbf F$]{Riemann-Silberstein-Weber vector}
\item[$\varepsilon$]{Order parameter}
\item[$\omega$]{Characteristic frequency} 
\item[${\mathcal O}$]{Big-O asymptotic notation}
\item[$\chi$]{Electric susceptibility}
\item[${\mathbb E}$]{Euclidean space}
\end{description}

%% file: intro.tex
\section{Introduction}
\label{sec:intro}

While major advances in the computational power of conventional computers have contributed to our understanding
of laboratory and space plasmas, the potential impact of quantum information science on plasma physics is especially
compelling and worthy of careful examination. In theory, quantum computers could perform some
numerical simulations exponentially faster than classical computers, which is cause for much optimism.
However, most plasmas of interest
do not exhibit quantum-like behavior. The constituent particles can be treated as classical point particles
since, as next illustrated, the interparticle distance in a plasma far exceeds the de Broglie wavelength of an electron.

In an ionized gas, the average interparticle distance is,\footnote{We will be using the SI (MKS) system of units.}
\begin{equation}
\langle \lambda \rangle_{ip} \approx \left( \frac{6}{\pi n} \right)^{1/3},
\end{equation}
where $n$ is the number density of particles. For densities ranging from $10^3$ to $10^{21}$ particles per cubic meter,
$\langle \lambda \rangle_{ip}$ is approximately between $0.1$ m and $10^{-7}$ m.
In contrast, the average de Broglie wavelength of an electron in an ideal gas is,
\begin{equation}
\langle \lambda \rangle_{dB}\approx \frac{h}{\sqrt{2 k_B T_e m_e}},
\end{equation}
where $h$ is the Planck constant, $k_B$ is the Boltzmann constant, $T_e$ is the electron temperature in Kelvin,
and $m_e$ is the mass of an electron.
For temperatures ranging from $10^4$ K ($\approx 1$ eV) to $2 \times 10^8$ K ($\approx 20$ keV),
$\langle \lambda \rangle_{dB}$ is approximately between $10^{-9}$ m and $10^{-13}$ m. 
It is evident that $\langle \lambda \rangle_{ip} \gg \langle \lambda \rangle_{dB}$ over a wide span of parameters.

Another length scale of significance is the electron Debye length,
\begin{equation}
\lambda_D =  \left( \frac{\epsilon_0}{e^2 n_e} k_B T_e \right)^{1/2},
\end{equation}
where $\epsilon_0$ is the permeability of vacuum, $e$ is the electron charge, and $n_e$ is the electron density.
The Debye length is a measure of the distance over which the electrostatic field of a stationary electron
is shielded. In three dimensional space, effective shielding requires many ions inside a Debye sphere (radius $\lambda_D$) centered at the location
of an electron; thus, $\lambda_D \gg \langle \lambda \rangle_{ip}$.\footnote{This inequality also implies that the electrostatic potential energy of an electron is
small compared to its thermal energy.} In order to avoid studying plasma behavior on a basis of point particles, we assume that 
$\langle \lambda \rangle_{ip} \ll \lambda_D \ll L_S$ where $L_S$ is the system size. These inequalities are satisfied by a wide variety of laboratory
and space plasmas. Consequently, we can transition from a single particle to a collective particle description of such plasmas. 
The collective behavior is well described by a statistical formulation in terms of particle distribution functions.  Alongside the kinetic
representation, we  can also develop a fluid description of a plasma by taking a hierarchy of momentum space moments of the distribution function. 

Based on the information provided thus far, one can rightfully question the relevance of quantum computers for exploring physical properties
of a classical plasma. This is where quantum information science becomes pertinent as it provides us an opportunity
to take on a broader, cross-disciplinary approach to developing algorithms for computational studies of plasma physics.
These algorithms can be implemented and tested on existing supercomputers and, subsequently, primed for quantum computers
while we await their general availability. In the meantime, there are several impediments to overcome before quantum computers can reliably
simulate practical topics in classical physics. For example, quantum bits (qubits) are sensitive to noise within their environment. The ensuing errors can accumulate,
leading to dubious results. Thus, development of error correction techniques for quantum computers is important.
We estimate that  typical simulations of plasma behavior will require quantum computers with thousands of qubits while maintaining
coherence over time scales entailing a few thousand 
logical quantum operations. Beside overcoming the technology challenges, we still have to
make advances in developing an interface for efficient interaction between the hardware and user-written software.

In spite of the obstacles mentioned above, the underlying principles that govern quantum computing provide a different perspective on
developing algorithms for numerical computations of diverse phenomena in plasmas. While deferring a discussion on the basic requisites of a quantum
computer to the next paragraph, we reflect on a particular topic of interest, namely, the propagation of electromagnetic waves in plasmas.
The occurrence of waves is a manifestation of the collective particle, or fluid-like, behavior of a plasma.
The waves are a consequence of instabilities excited by readily tappable sources of free energy inherent in a plasma. Remote observations, identifying different
characteristics of the waves, provide useful information about the plasma itself, for example, electromagnetic emissions from the solar corona furnish details
of the composition of coronal plasma. 
Furthermore, we can artificially generate electromagnetic waves by oscillating charges, and use them for a myriad of well-known applications, as in
communication, medical imaging, remote sensing, and controlling the plasma environment inside a nuclear fusion reactor.
The propagation of electromagnetic waves is described by Maxwell equations in which a plasma
is characterized as a dielectric medium with its polarization density included in the electric displacement field. We will assume that
the polarization density and, as a consequence, the electric displacement field is a linear function of the electric field amplitude. Physically, the
linear relation, a core assumption of {\it linear response theory},  implies that the external electromagnetic field does not change the dielectric properties of a plasma.
Linear response theory is widely used for analyzing the propagation, scattering, and damping
of electromagnetic waves in a variety of laboratory and space plasmas.
Consequently, Maxwell equations are linearly dependent on the electromagnetic fields and well suited
for implementing  in quantum computers.

The operations of a quantum computer are based on the axioms of
quantum physics. The prevailing theory of quantum mechanics is structured around linear operators acting on a state function whose
time evolution is prescribed by the Schr\"odinger equation.
Thus, for implementing in a quantum computer, the classical Maxwell equations have to be converted
to a form which is similar to the Schr\"odinger equation. The recasting has to preserve the
physics, in particular, the conservation principles, of the original Maxwell equations. The tools of linear algebra
provide the basic foundation for a systematic conversion of a classical system to a quantum-like structure.
Following such a conversion, the reformulated Maxwell equations are amenable to quantum computing.

This chapter is organized accordingly as follows: Section \ref{sec:tensor} covers topics in linear algebra which are necessary
to repurpose a classical set of equations for implementation in a  quantum computer. This section elucidates the symbiotic,
and salient, relation between vector spaces and dual spaces that is essential in classical and quantum physics. Furthermore, the
section includes a discussion on covariant and contravariant vectors, tensors as tensor product of vectors, the transformation 
between vectors and covectors, and Hilbert vector spaces. In Section \ref{braket}, we introduce Dirac's notation and
define operators in Hilbert space. The quantum postulates that govern quantum computers are enunciated in Section \ref{pqm},
which is followed by Section \ref{sec:basicqc} on the basics of quantum computing. Section \ref{sec:maxwellco} details the covariant
form of Maxwell equations and, in Section \ref{sec:urmev}, their subsequent conversion to a unitary set of equations for electromagnetic wave
propagation in vacuum. A quantum lattice algorithm for wave propagation in a dielectric medium is mapped out in
Section \ref{sec:qla}. We conclude the chapter with some final thoughts in Section \ref{sec:conc}.

A concise narrative on Euclidean, Cartesian, and non-Euclidean spaces is given in the Appendix, Section \ref{appendix}.
There is substantial literature available on the topics we have covered in this chapter, and the Bibliography
Section \ref{sec:supp} contains selected references we have found useful in our studies.

%% file: tensor1.tex
\section{An Introduction to Vector Spaces, Tensors, and  Hilbert Spaces}
\label{sec:tensor}

\epigraph{L'alg\`ebre est g\'en\'ereuse, elle donne souvent plus qu'on lui
demande. \\ 
Algebra is generous; she often gives more than is asked of her.}{Jean le Rond D’Alembert $\quad \quad$ \\
Quote from E. Kasner in $\quad\quad$ \\
Bull. Am. Math. Soc. {\bf 11}, 283 (1905)}

This section is on some basic aspects of linear algebra which will be useful in
formulating theoretical models and computational algorithms suitable for quantum computers.

\subsection{\textbf{\textit{Fields}}}
\label{sub:fld}

In linear algebra, a {\it field} is the basic foundation on which we build vector spaces and tensor
spaces -- concepts which we will need later to formulate classical electromagnetism within the
framework of a quantum description.

A field, denoted by a triplet $\left( {\mathcal F}, +, \bcdot \right) $, is a non-empty set ${\mathcal F}$ consisting of
well-defined and distinct elements along with binary operations $+$ (addition) and $\bcdot$ (multiplication),
\begin{align}
  &\begin{aligned}
    + : {\mathcal F} \times {\mathcal F} & \to  {\mathcal F} \\
    \left( a , b \right) & \mapsto   a + b \quad \quad \forall \ a, b  \in {\mathcal F}
  \end{aligned}\\[8pt]
  &\begin{aligned}
    \bcdot : {\mathcal F} \times {\mathcal F} & \to  {\mathcal F} \\
    \left( a, b \right) & \mapsto  a \cdot b \quad \quad \forall \ a, b \in {\mathcal F}
  \end{aligned}
\end{align}
The two operations have to satisfy certain axioms.\footnote{The {\it Cartesian product} ${\mathcal F} \times {\mathcal F}$ is a set
of all ordered pairs $\left( a , b \right)$ where $a$ belongs to the first field and $b$ to the second field.}
For all ${a, b, c} \in {\mathcal F}$, the axioms for $+$ are,
\TabPositions{8cm}
\begin{itemize}
\item $a + b \in {\mathcal F}$ \tab (closure)  
\item $a + b = b + a$ \tab (commutative)  
\item $(a + b) + c = a + (b + c)$ \tab (associative)  
\item $\exists$ an element $\left\{ 0 \in {\mathcal F} \ \big\vert \ a + 0 = 0 + a = a \right\} $ \tab (identity)  
\item for any $a$ $\exists$ $\left\{ b \ \big\vert \ a + b = b + a = 0 \right\}$ \tab (inverse)  
\end{itemize} 
while the axioms for $\bcdot$ are,
\begin{itemize}
\item $a \bcdot b \in {\mathcal F}$ \tab (closure) 
\item $a \bcdot  b = b \bcdot a$ \tab (commutative) 
\item $(a \bcdot b) \bcdot c = a \bcdot (b \bcdot c)$ \tab (associative) 
\item $\exists$ an element $\left\{ 1 \in {\mathcal F}\backslash \{0\} \ \big\vert \ a \bcdot 1 = 1 \bcdot a = a \right\} $ \tab (identity)  
\item for any $a \in {\mathcal F} \backslash \{0\}$ $\exists$ $\left\{ b \ \big\vert \ a \bcdot b = b \bcdot a = 1 \right\}$ \tab (inverse)  
\end{itemize}
Additionally, there are two distributive properties that combine $+$ and $\bcdot$,
\begin{itemize}
\item $ a \bcdot \left( b + c \right) = a \bcdot b + a \bcdot c $ \tab (left distributive)
\item $ \left( a + b \right) \bcdot c = a \bcdot c + b \bcdot c $\tab (right distributive)
\end{itemize}
In linear (abstract) algebra terminology, $\left( {\mathcal F}, \ + \right)$ and,
$\left( {\mathcal F} \backslash \{0\}, \ \bcdot \right)$ are abelian groups, while $\left( {\mathcal F}, + , \bcdot \right)$ is a
commutative ring; $\left( {\mathcal F} \backslash \{0\} \right)$ indicates
the set $\mathcal F$ without the additive identity element. 

For a field ${\mathcal F} = {\mathbb R}$ which is a set of real numbers,  
operations $+$ and  $\bcdot$ represent the conventional addition and multiplication operations, respectively. The additive inverse
is $-a$ for all $a \in {\mathcal F}$, and the multiplicative inverse is $a^{-1}$ for all $a \in \left( {\mathcal F} \backslash \{0\} \right)$.

For a field ${\mathcal F} = {\mathbb C}$ which is a set of complex numbers, we write each element as an ordered pair,
\begin{equation}
{\mathbb C} = \left\{ (a, b) \ \big\vert \ a, b \in {\mathbb R} \right\}.
\end{equation}
By an ordered pair we mean that $(a, b) \ne (b, a)$ unless $a = b$. The two binary operations are,
\begin{align}
    + : {\mathbb C} \times {\mathbb C} & \to  {\mathbb C} \quad \Longrightarrow  \quad
         \bigl( \left( a, b \right), \left( c, d \right) \bigr) & \mapsto & \ \  \bigl( \left( a + c \right), \left( b + d \right) \bigr), \\
    \bcdot : {\mathbb C} \times {\mathbb C} & \to  {\mathbb C}  \quad \Longrightarrow \quad
    \bigl( \left( a, b \right), \left( c , d \right) \bigr) & \mapsto & \  \ \bigl( \left( ac - bd \right), \left( ad + bc \right) \bigr),
\end{align}
with $c, d \in {\mathbb R}$.

Henceforth, we will refer to elements in the field $\mathcal F$ as scalars. Furthermore, unless specified otherwise, 
${\mathcal F} = \mathbb R$ or ${\mathcal F} = \mathbb C$ (${\mathbb R} \subset \mathbb C$).

\subsection{\textbf{\textit{Linear Vector Spaces}}}
\label{sub:vs}

A {\it vector space} over a field ${\mathcal F}$ denoted by a triplet $\left( {\bm V}, +, \bcdot \right)$, comprising of
a non-empty set ${\bm V}$ with two binary operations, vector addition and scalar multiplication, is defined as,\footnote{The Cartesian product 
${\bm V} \times {\bm V}$ is a set
of all ordered pairs of {\it vectors} $\left( {\mathbf u} , {\mathbf v} \right)$ 
where ${\mathbf u}$ belongs to the first vector space and ${\mathbf v}$ to the second vector space.}$^,$\footnote{Even though we use the same 
symbols for binary operations for fields and vector spaces, their properties are different. However, the similarity in notation should not lead to any confusion.}
\begin{align}
  &\begin{aligned}
    + : {\bm V} \times {\bm V} & \to  {\bm V} \\
    \left( {\mathbf u}, {\mathbf v}  \right) & \mapsto   {\mathbf u} + {\mathbf v}  \quad \quad \forall \ {\mathbf u}, {\mathbf v} \in {\bm V} \\
  \end{aligned}\\[8pt]
  &\begin{aligned}
    \cdot : {\mathcal F} \times {\bm V}  & \to  {\bm V} \\
    \left( a, {\mathbf u} \right) & \mapsto  a \cdot {\mathbf u} \quad \quad \forall \ a \in {\mathcal F} \ {\rm and} \ \forall \ {\mathbf u} \in {\bm V}
  \end{aligned}
\end{align} 
The binary operations have to satisfy the following axioms,
\TabPositions{7.8cm}
\begin{itemize}
\item ${\mathbf u } + {\mathbf v} = {\mathbf v } + {\mathbf u} \in {\bm V}$ \tab (commutative)
\item ${\mathbf u } + \left( {\mathbf v} + {\mathbf w} \right) = \left( {\mathbf u } + {\mathbf v} \right)  + {\mathbf w}$ \tab (associative)
\item  $\exists$ a vector $\left\{ {\mathbf 0} \in {\bm V} \ \middle| \ {\mathbf u} + {\mathbf 0} = {\mathbf 0} + {\mathbf u} = {\mathbf u} \right\} $ \tab (null vector) 
\item  $\exists$ a vector $\left\{ {\mathbf v} \in {\bm V} \ \middle| \ {\mathbf u} +  {\mathbf v} 
 = {\mathbf 0} \right\}$ \tab (inverse) 
\item $a \cdot \left( b \cdot  {\mathbf u} \right) = a \cdot  b \cdot \left( {\mathbf u} \right)$ \tab (scalar multiplication)
\item $ a \cdot \left( {\mathbf u}+ {\mathbf v} \right) = a \cdot {\mathbf u} + a \cdot {\mathbf v}$ and 
$\left( a + b \right) \cdot {\mathbf u} = a \cdot {\mathbf u} + b \cdot {\mathbf u}$ \tab
(distributive)
\item $\exists$ $\left\{ 1 \in {\mathcal F} \ \middle| \ 1 \cdot {\mathbf u} = {\mathbf u} \right\}$ \tab (multiplicative identity)
\end{itemize}

While this is a purely abstract definition, it is easy to connect it to our conventional understanding of vectors in the Euclidean plane ${\mathbb R}^2$;
that is, let ${\bm V} = \mathbb R^2$.
With respect to a prescribed origin and Cartesian coordinate system, we represent each vector by a $2 \times 1$ column vector,
\begin{equation}
{\mathbf u} = 
\begin{bmatrix}
u_x \\
u_y 
\end{bmatrix}, 
\quad
{\mathbf v} = 
\begin{bmatrix}
v_x \\
v_y 
\end{bmatrix}, 
\quad
{\mathbf 0} = 
\begin{bmatrix}
0 \\
0 
\end{bmatrix},
\end{equation}
where the upper (lower) entry is the $x$ ($y$) component of the vector. It is straightforward to
realize the axioms stated above with $ a, b \in {\mathcal F} =  {\mathbb R}$.  The scalar multiplication should not
be confused with the vector dot product in conventional treatment of vectors.

The importance of an abstract definition of a vector space is that we can work in any number of dimensions
with any representation of the field $\mathcal F$ and of the vector space ${\bm V}$, as long as the indicated axioms
are satisfied.

\subsection{\textbf{\textit{Finite Dimensional Vector Spaces}}}

In this subsection, we will assume that ${\bm V}$ is a non-empty vector space over a field ${\mathcal F}$.

\subsubsection{Subspace}
\label{sub:subspaceA}

If ${\bm W}$ is a subset of vectors taken from ${\bm V}$,  then ${\bm W}$ 
is also a vector space if and only if, for any ${\mathbf u}, \, {\mathbf v} \in {\bm W}$ and $a \in {\mathcal F}$,
\begin{itemize}
\item ${\mathbf u} + {\mathbf v} \in {\bm W}$
\item $a {\mathbf u} \in {\bm W}$.
\end{itemize}
We refer to ${\bm W}$ as a {\it subspace} of ${\bm V}$; ${\bm W} \subset {\bm V}$.

\subsubsection{Linear Independence}

Consider a subset of $n \in {\mathbb N}$ vectors from ${\bm V}$,
\begin{equation}
{\bm S} = \Bigl\{ {\mathbf e}_1, {\mathbf e}_2, \dots, {\mathbf e}_n \ \Bigr\vert \ {\mathbf e}_i \in {\bm V}, \ \ i = 1, 2, \dots, n \Bigr\}.
\end{equation}
The vectors in ${\bm S}$ are {\it linearly independent} if and only if the equation,
\begin{equation}
a_1 {\mathbf e}_1 + a_2 {\mathbf e}_2 + \dots \dots + a_n {\mathbf e}_ n \ = \ 0,  \quad \quad \left\{ a_i \in {\mathcal F} \mid i = 1, 2, \dots , n \right\},
\end{equation}
has only the trivial solution $a_i = 0$, for all $i = 1, 2, \dots, n$. Otherwise, the vectors ${\mathbf e}_i$ are linearly dependent. 

\subsubsection{Span, Coordinates, and Basis Set}

Let,
\begin{equation}
{\mathcal B} = \left\{ {\mathbf e}_1, {\mathbf e}_2, \dots \dots, {\mathbf e}_n  \ \Big\vert \ {\mathbf e}_i \in {\bm V}, \ n \in {\mathbb N} \right\}, 
\label{basis1}
\end{equation}
be a set of ordered linearly independent vectors.\footnote{We assume that ${\mathcal B}$ is a {\it maximal} linearly independent set --
it is not a subset of any other linearly independent set.}
We define the {\it span} of ${\mathcal B}$ to be a set of all linear combinations 
of $\{  {\mathbf e}_1, {\mathbf e}_2, \dots, {\mathbf e}_n \} $,
\begin{equation}
{\rm span} \left( {\mathcal B} \right) = \Bigl\{ x^1 {\mathbf e}_1 +  x^2 {\mathbf e}_2 + \dots + x^n {\mathbf e}_n \ \Big\vert \ n \in {\mathbb N}, \ 
x^i \in {\mathcal F}, \ {\mathbf e_i} \in {\bm V} \Bigr\}. \label{basis1AA}
\end{equation}
While ${\mathcal B}$ is a subset of ${\bm V}$, ${\rm span} ( {\mathcal B} )$ is a vector subspace; the conditions
stated in Subsection \ref{sub:subspaceA} are certainly satisfied.
In \eqref{basis1AA}, 
\begin{equation}
 X = \left\{ x^1, x^2, \dots \dots, x^n \ \Big\vert \ x^i \in {\mathcal F}, \ i = 1, 2, \dots, n \right\},
\end{equation}
is a set of coefficients 
referred to as {\it coordinates}. The superscripts are indices and not exponents.\footnote{An explanation of this notation is given 
later in the section on contravariant and covariant vectors.}
We refer to the set ${\mathcal B}$ as the {\it basis} set for the vector subspace ${\rm span} \left( {\mathcal B} \right)$.  The {\it dimension}
of ${\rm span} \left( {\mathcal B} \right)$ is defined to be $n$ -- the number of independent basis vectors that span the subspace.
The abbreviated expression is ${\rm dim} \left( {\rm span} \left( {\mathcal B} \right) \right) = n$.
In the $n$-dimensional subspace, each element of the basis set is a $n$-tuple column vector,
\begin{equation}
{\mathbf e}_i = \begin{bmatrix}  e_{i1} & e_{i2} & \dots & e_{in} \end{bmatrix}^{\rm T},\quad \quad (e_{ij} \in {\mathcal F} \ {\rm for} \ i, j = 1, 2, \dots n),
\end{equation}
where the superscript ${\rm T}$ indicates the transpose of the row vector.

If ${\rm span} \left( {\mathcal B} \right) = {\bm V}$, that is, the span of ${\mathcal B}$ is the entire vector space ${\bm V}$, then
the elements of ${\mathcal B}$ are the {\it basis vectors} for the vector space ${\bm V}$. The dimension of ${\bm V}$ is 
${\rm dim}  \left( {\bm V} \right) = n$, and any vector ${\mathbf u} \in {\bm V}$ can be written uniquely as,
\begin{equation}
{\mathbf u} = u^1 {\mathbf e}_1 + u^2 {\mathbf e}_2 + \dots \dots + u^n {\mathbf e}_n  = \sum_{i=1}^n \ u^i {\mathbf e}_i, \label{basis2}
\end{equation}
where $u^i$ $\left( i = 1, 2, \dots, n \right)$ are the coordinates of ${\mathbf u}$ corresponding to the basis set ${\mathcal B}$.
Each vector in ${\bm V}$ has a unique set of coordinates with respect to the basis set ${\mathcal B}$, and is an ordered set over a $n$-dimensional
field ${\mathcal F}^n = \underbrace{ {\mathcal F} \times {\mathcal F} \dots \times {\mathcal F}}_{n \,\, {\rm times}} $.
For a  finite dimensional vector space there are an infinite number of basis sets; however, all of them have the same dimension as that of the
vector space. Consequently, for a given physical situation, we can choose a basis set that is convenient for mathematical analysis.

\subsection{\textbf{\textit{Linear Maps, Linear Operators, Linear Functionals, and Dual Spaces}}}

\subsubsection{Linear maps and linear operators}
\label{sub2:linearmap}

For two vector spaces ${\bm V}$ and ${\bm W}$ over the field ${\mathcal F}$,
\begin{equation} 
f : {\bm V} \to {\bm W} \quad \Rightarrow \quad {\mathbf v} \mapsto  f \left( {\mathbf v} \right), \quad {\mathbf v} \in {\bm V}, \ f( {\mathbf v} ) \in {\bm W},  
\end{equation} 
is a {\it linear map} or {\it linear transformation} if and only if,  
\begin{equation} 
f \left( {\mathbf u} + {\mathbf v} \right) = f \left( {\mathbf u} \right) + f \left(  {\mathbf v} \right), \quad {\rm and} \quad f \left( a {\mathbf v} \right) =  
a f \left( {\mathbf v} \right),  
\end{equation} 
for all ${\mathbf u}, {\mathbf v}  \in {\bm V}$ and $a \in {\mathcal F}$.  
The set of all such linear maps is ${\mathcal L} \left( {\bm V}, {\bm W} \right)$.  
 
The linear map $f$ is {\it injective} or {\it one-to-one} if $ f \left( {\mathbf u} \right) = f \left( {\mathbf v} \right) $ implies 
$ {\mathbf u} = {\mathbf v}$. The map is {\it surjective} or {\it onto} if for all $ {\mathbf w} \in {\bm W}$ there exists 
$ {\mathbf v} \in {\bm V} $ such that $ f \left(  {\mathbf v} \right) =  {\mathbf w} $. The mapping is both injective and surjective,  
or {\it bijective}, if there exists a unique $ {\mathbf v} $ such that $ f \left(  {\mathbf v} \right) =  {\mathbf w} $ for all $ {\mathbf w} \in {\bm W} $.  

A linear transformation, 
\begin{equation}
f: {\bm V} \to {\bm V} \Rightarrow  {\mathbf v} \mapsto  f \left( {\mathbf v} \right), \quad {\mathbf v}, f \left( {\mathbf v} \right) \in {\bm V}, \label{linop1}
\end{equation}
is called a {\it linear operator}, or an {\it endomorphism} on ${\bm V}$. If in the basis set \eqref{basis1},
${\mathbf v} = v^1 {\mathbf e}_1 + v^2 {\mathbf e}_2 + \dots + v^n {\mathbf e}_n$,
where $v^1, v^2, \dots , v^n \in {\mathcal F}$, then,
\begin{equation}
f \left( {\mathbf v} \right) = f \left( \sum\limits_{i=1}^n v^i {\mathbf e}_i \right) = \sum\limits_{i=1}^n v^i f \left( {\mathbf e}_i \right)
= \sum\limits_{i=1}^n v^i f_i. \label{linop2}
\end{equation}
Here we have used the linearity of the operator $f$ and made the substittuion $f_i = f \left( {\mathbf e}_i \right)$. 
Since $f_i \in {\bm V}$, it can also be expanded in terms of the basis set \eqref{basis1},
$ f_i = f_i^1 {\mathbf e}_1 + \dots + f_i^n {\mathbf e}_n$ where $f_i^j \in {\mathcal F}$ for $ j = 1, \dots , n $. Subsequently,
\begin{equation}
f \left( {\mathbf v} \right)  = \sum\limits_{i, j = 1}^n \ v^i f_i^j {\mathbf e}_j, \label{linop3}
\end{equation}
is an expansion of the action of a linear operator on a vector in the prescribed basis set.

\subsubsection{Linear Functional}

A mapping,
\begin{equation}
    f :\ {\bm V}  \ \to \  {\mathcal F} \quad \Longrightarrow \quad 
     {\mathbf u} \ \mapsto \ f ( {\mathbf u} ),  \quad \quad f ( {\mathbf u} ) \in {\mathcal F} \ \forall \ {\mathbf u} \in {\bm V}, \label{mapfunc}
\end{equation}
is a {\it linear functional} on ${\bm V}$ over ${\mathcal F}$ if and only if,
\begin{equation}
f \left( a {\mathbf u} + b {\mathbf v} \right) = a f ( {\mathbf u} )  + b f ( {\mathbf v} ), \quad \forall \ {\mathbf u}, {\mathbf v} \in {\bm V}, \ \ 
a, b \in {\mathcal F}.
\end{equation}

\subsubsection{Dual Space}

Let ${\bm V}^\ast$ be the set of all linear functionals from vector space ${\bm V}$ to ${\mathcal F}$. If, for all 
$ {\mathfrak f} , {\mathfrak g}  \in {\bm V}^*$,
${\mathbf u} \in {\bm V}$, and $a \in {\mathcal F}$, we satisfy,
\begin{equation}
\left( {\mathfrak f} + {\mathfrak g} \right) \left( {\mathbf u} \right) =  {\mathfrak f} \left( {\mathbf u} \right)  + 
{\mathfrak g} \left( {\mathbf u} \right), \quad {\rm and} \ \
\left( a {\mathfrak f} \right) \left( {\mathbf u} \right)= a \left( {\mathfrak f} \left( {\mathbf u} \right) \right), \label{dual1}
\end{equation}
then ${\bm V}^*$ is a vector space -- it is the {\it dual space} of ${\bm V}$. For the basis set $\mathcal B$ in \eqref{basis1},
consider a set of linear functionals,
\begin{equation}
{\mathcal B}^* = \left\{ {\mathbf f}^1, {\mathbf f} ^2, \dots, {\mathbf f} ^n \ \Big\vert \ {\mathbf f} ^i \in {\bm V}^* \ {\rm for} \ i = 1, 2, \dots, n \right\}, \label{dual2}
\end{equation}
where ${\mathbf f} ^i$ has the following functional property,
\begin{equation}
{\mathbf f}^i \left( {\mathbf e}_j \right) = \delta^i_j = \begin{cases} 0, & \ {\rm if}\ i \ne j \\ 1, & \ {\rm if}\ i = j,  \quad \forall \ i, j = 1, 2, \dots, n \end{cases}
\label{dual3}
\end{equation}
with ${\mathbf e}_j \in {\mathcal B}$, and  $\delta^i_j$ being the Kronecker delta. 
Subsequently, it is straightforward to prove that ${\mathcal B}^*$ is a basis set for ${\bm V}^*$ -- its elements are linearly independent and it
spans ${\bm V}^*$. ${\mathcal B}^*$ is the {\it dual basis} for ${\bm V}$ with ${\rm dim} ({\bm V}^*) = {\rm dim} ({\bm V})$.
The elements in ${\bm V}^*$ are named {\it covectors}.
Applying ${\mathbf f} ^i$ to \eqref{basis2}, we obtain,
\begin{equation}
{\mathbf f} ^i \left( {\mathbf u} \right)  =  {\mathbf f} ^i \left( \sum_{j=1}^{n} \ u^j {\mathbf e}_j\right)
 =  \sum_{j=1}^{n} \ u^j \, {\mathbf f} ^i \left( {\mathbf e}_j \right)
 =  \sum_{j=1}^{n} \ u^j \, \delta^i_j  = u^i. \label{dual4}
\end{equation}
Thus, the functional ${\mathbf f} ^i$ maps ${\mathbf u}$ into its $i$-th coordinate. Consequently,
\begin{equation}
{\mathbf u} = \sum\limits_{i=1}^n {\mathbf f} ^i \left( {\mathbf u} \right) \,  {\mathbf e}_i ,
\label{dual5}
\end{equation}
and, any linear functional ${\mathfrak g} \in {\bm V}^*$ can be expressed in terms of the dual basis, 
\begin{equation}
{\mathfrak g} = \sum\limits_{i=1}^n {\mathfrak g}  \left( {\mathbf e}_i \right) \,  {\mathbf f}^i .
\label{dual6}
\end{equation}
The equality in \eqref{dual6} can be proven by applying ${\mathfrak g}$, on the left hand side, to any basis vector ${\mathbf e}_i$.
Alternatively, we can directly expand ${\mathfrak g}$ in the dual basis,
\begin{equation}
{\mathfrak g} = \sum\limits_{i =1}^n \ \xi_i {\mathbf f}^i,
\quad \quad \left\{\xi_1, \xi_2, \dots , \xi_n \right\} \in {\mathcal F},
\label{dual7}
\end{equation}
and, upon applying this expression to \eqref{basis2}, obtain,
\begin{equation}
{\mathfrak g} \left( {\mathbf u}\right) = \sum_{i=1}^n \ \xi_i u^i. \label{dual8}
\end{equation}
This form is akin to the dot product of two vectors ${\mathfrak g}$ and ${\mathbf u}$.

\subsubsection{Contravariant and Covariant Vectors}
\label{subsub:contracov}

In the vector space ${\bm V}$ with ${\rm dim} \left( V \right) = n$, consider two different sets of basis vectors,
\begin{equation}
{\mathcal B} = \left\{ {\mathbf e}_1, {\mathbf e}_2, \dots , {\mathbf e}_n \right\}, \quad \quad  
\widetilde{\mathcal B} = \left\{\tilde{\mathbf e}_1, \tilde{\mathbf e}_2, \dots , \tilde{\mathbf e}_n \right\}.  \label{cc1}
\end{equation} 
We can express each element in $\widetilde{\mathcal B}$ as a linear sum of the basis vectors in ${\mathcal B}$,
\begin{equation}
\tilde{\mathbf e}_i= \sum_{j=1}^n \ a_i^j {\mathbf e}_j, \quad \quad a_i^j \in {\mathcal F},  \quad \quad  i, j = 1, 2, \dots, n. 
\label{cc2}
\end{equation}
This equation can be written in a compact form,
\begin{equation}
\widetilde{\bm E} = {\bm A} \ {\bm E}, \label{cc3}
\end{equation}
where 
\begin{equation}
\widetilde{\bm E} = \begin{bmatrix} \tilde{\mathbf e}_1 \\[6pt] \tilde{\mathbf e}_2 \\[6pt] \vdots \\[6pt] \tilde{\mathbf e}_n \end{bmatrix}, \quad \quad 
{\bm E} = \begin{bmatrix} {\mathbf e}_1 \\[6pt] {\mathbf e}_2 \\[6pt] \vdots \\[6pt] {\mathbf e}_n \end{bmatrix}, \quad \quad 
{\bm A} = \begin{bmatrix} a_1^1 & a_1^2 & \dots & a_1^n \\[6pt]
a_2^1 & a_2^2 & \dots & a_2^n \\[6pt]
\vdots & \vdots & \ddots & \vdots \\[6pt]
a_n^1 & a_n^2 & \dots & a_n^n 
\end{bmatrix}. \label{cc4}
\end{equation}
 Let ${\mathcal B}^*$ and $\widetilde{\mathcal B}^*$ be two different sets of functional basis vectors belonging to the dual vector space ${\bm V}^*$,
\begin{equation}
{\mathcal B}^* = \left\{ {\mathbf f}^1, {\mathbf f}^2, \dots , {\mathbf f}^n \right\}, \quad \quad 
\widetilde{\mathcal B}^* = \left\{\tilde{\mathbf f}^1, \tilde{\mathbf f}^2, \dots , \tilde{\mathbf f}^n \right\}, \label{cc5}
\end{equation}
where, following \eqref{dual3}, 
\begin{equation}
  {\mathbf f}^i\left( {\mathbf e}_j \right) = \delta^i_j, \quad\quad 
\tilde{\mathbf f}^i\left( \tilde{\mathbf e}_j \right) = \delta^i_j.  \label{cc6}
\end{equation}
The two dual functional basis sets are related,
\begin{equation}
\tilde{\mathbf f}^i= \sum_{j=1}^n \ b_j^i {\mathbf f}^j, \quad \quad b_j^i \in {\mathcal F},  \quad \quad  i, j = 1, 2, \dots, n,
\label{cc7}
\end{equation}
which takes on the compact form,
\begin{equation}
\widetilde{\bm F} = {\bm B} \ {\bm F}, \label{cc8}
\end{equation}
where,
\begin{equation}
\widetilde{\bm F} = \begin{bmatrix} \tilde{\mathbf f}^1 \\[6pt] \tilde{\mathbf f}^2 \\[6pt] \vdots \\[6pt] \tilde{\mathbf f}^n \end{bmatrix}, \quad \quad 
{\bm F} = \begin{bmatrix} {\mathbf f}^1 \\[6pt] {\mathbf f}^2 \\[6pt] \vdots \\[6pt] {\mathbf f}^n \end{bmatrix}, \quad \quad 
{\bm B} = \begin{bmatrix} b_1^1 & b_2^1 & \dots & b_n^1 \\[6pt]
b_1^2 & b_2^2 & \dots & b_n^2 \\[6pt]
\vdots & \vdots & \ddots & \vdots \\[6pt]
b_1^n & b_2^n & \dots & b_n^n 
\end{bmatrix}. \label{cc9}
\end{equation}
From \eqref{cc2} and \eqref{cc7},
\begin{equation}
\begin{alignedat}{2}
\tilde{\mathbf f}^i \left( \tilde{\mathbf e}_k \right)  = & \sum_{j=1}^n \  b^i_j \,  {\mathbf f}^j \left( \tilde{\mathbf e}_k \right)
= &&  \sum_{j=1}^n \ b^i_j  \, {\mathbf f}^j \left( \sum_{l=1}^n  \ a^l_k {\mathbf e}_l \right) 
= \sum_{j=1}^n \sum_{l=1}^n  \ b^i_j  a^l_k \, {\mathbf f}^j \left( {\mathbf e}_l \right) \\
= &  \sum_{j=1}^n \ \sum_{l=1}^n  \  b^i_j  a^l_k \delta^j_l 
= &&  \sum_{l=1}^n \  b^i_l  a^l_k. \label{cc10}
\end{alignedat}
\end{equation}
Combining \eqref{cc10} with \eqref{cc6}, and using \eqref{cc4} and \eqref{cc9}, we find,
\begin{equation}
{\bm A}\ {\bm B}^{\rm T} \ = \ {\mathcal I}_n \ \quad \Rightarrow \quad {\bm B}^{\rm T} \ = \ {\bm A}^{-1}, \label{cc11}
\end{equation}
where ${\mathcal I}_n$ is a $n \times n$ identity matrix.
This relationship between ${\bm A}$ and ${\bm B}$ is important in understanding the difference between covariant
and contravariant vectors.

An element ${\mathfrak g} \in {\bm V}^*$ has the following representation in the two sets of basis vectors ${\mathcal B}^*$ 
and $\widetilde{\mathcal B}^*$ in the dual space,
\begin{equation}
{\mathfrak g} = \begin{bmatrix} {\mathbf f}^1 & {\mathbf f}^2 & \dots & {\mathbf f}^n \end{bmatrix} 
\begin{bmatrix} g_1 \\ g_2 \\ \vdots \\ g_n \end{bmatrix}, 
\quad\quad 
{\mathfrak g} = \begin{bmatrix} \tilde{\mathbf f}^1 & \tilde{\mathbf f}^2 & \dots & \tilde{\mathbf f}^n \end{bmatrix} 
\begin{bmatrix} {\tilde g}_1 \\ {\tilde g}_2 \\ \vdots \\ {\tilde g}_n \end{bmatrix}, \label{cc12}
\end{equation}
where $g_i \in {\mathcal F}$ and $\tilde{g}_i \in {\mathcal F}$ $\left( i = 1, 2, \dots , n \right)$ 
are the appropriate coordinates of ${\mathfrak g}$ in the corresponding basis sets.
Substituting \eqref{cc9} into the first expression in \eqref{cc12}, and using \eqref{cc11},
\begin{equation}
{\mathfrak g} = \begin{bmatrix} \tilde{\mathbf f}^1 & \tilde{\mathbf f}^2 & \dots & \tilde{\mathbf f}^n \end{bmatrix} \ \left( {\bm B}^{\rm T} \right)^{-1} \ 
\begin{bmatrix} {\tilde g}_1 \\ {\tilde g}_2 \\ \vdots \\ {\tilde g}_n \end{bmatrix}
= \begin{bmatrix} \tilde{\mathbf f}^1 & \tilde{\mathbf f}^2 & \dots & \tilde{\mathbf f}^n \end{bmatrix} \ {\bm A}
\begin{bmatrix} {g}_1 \\ {g}_2 \\ \vdots \\ {g}_n \end{bmatrix}. \label{cc13}
\end{equation}
Comparing with the second equation in \eqref{cc12}, we note that the coordinates transform in the same way as the 
basis vectors in \eqref{cc3},
\begin{equation}
\begin{bmatrix} {\tilde g}_1 \\ {\tilde g}_2 \\ \vdots \\ {\tilde g}_n \end{bmatrix}
\ = \ {\bm A} \ \begin{bmatrix} {g}_1 \\ {g}_2 \\ \vdots \\ {g}_n \end{bmatrix}.
\label{cc14}
\end{equation}
Consequently, vectors in the dual space ${\bm V}^*$ are referred to as {\it covariant} vectors -- their coordinates transform in the same
way as the basis vectors in ${\bm V}$.

For ${\mathbf u} \in {\bm V}$ having the following form in the two basis sets ${\mathcal B}$ and $\widetilde{\mathcal B}$,
\begin{equation}
{\mathbf u} = u^1 {\mathbf e}_1 + u^2 {\mathbf e}_2 + \dots + u^n {\mathbf e}_n,\quad \quad
{\mathbf u} = \tilde{u}^1 \tilde{\mathbf e}_1 + \tilde{u}^2 \tilde{\mathbf e}_2 + \dots + \tilde{u}^n \tilde{\mathbf e}_n, \label{cc15}
\end{equation}
we can similarly show that,
\begin{equation}
\begin{bmatrix} {\tilde u}^1 \\ {\tilde u}^2 \\ \vdots \\ {\tilde u}^n \end{bmatrix}
\ = \ \left( {\bm A}^{-1} \right)^{\rm T}  \ \begin{bmatrix} {u}^1 \\ {u}^2 \\ \vdots \\ {u}^n \end{bmatrix}. \label{cc16}
\end{equation}
The coordinates in ${\bm V}$ transform according to the reciprocal basis -- inverse of the way in which the basis vectors 
in ${\bm V}$ transform. Thus, vectors in ${\bm V}$ are {\it contravariant}. 

For a covariant vector we will use subscripts for the coordinates while using superscripts for the coordinate of a contravariant vector.
The covariant vectors are linear mappings on the space of contravariant vectors.

\subsection{\textbf{\textit{Direct Sum, Bilinear Maps, Tensor Product}}}

\subsubsection{Direct Sum}

A vector space ${\bm V}$ is a {\it direct sum} of its subspaces ${\bm U}$ and ${\bm W}$, ${\bm V} = {\bm U} \oplus {\bm W}$, if 
${\bm U} \cap {\bm V} = \left\{ {\bm 0} \right\}$ and 
for all ${\mathbf v} \in {\bm V}$ there is a unique ${\mathbf u} \in {\bm U}$ and ${\mathbf w} \in {\bm W}$ such that ${\mathbf v} = 
{\mathbf u} + {\mathbf w}$; ${\bm V}$, ${\bm U}$, and ${\bm W}$, are vector spaces over the same field ${\mathcal F}$. 
If ${\rm dim} ({\bm U}) = m$ with basis set $\left\{ {\mathbf e}_1, {\mathbf e}_2, \dots , {\mathbf e}_m\right\} $
and ${\rm dim} ({\bm W}) = n$ with basis set $\left\{ \tilde{\mathbf e}_1, \tilde{\mathbf e}_2, \dots , \tilde{\mathbf e}_n \right\}$, then 
the combined basis set of $m + n$ elements $\left\{ {\mathbf e}_1, {\mathbf e}_2, \dots , {\mathbf e}_m, 
\tilde{\mathbf e}_1, \tilde{\mathbf e}_2, \dots , \tilde{\mathbf e}_n \right\}$ spans ${\bm V} = {\bm U} \oplus {\bm W}$. If, 
\begin{equation}
{\mathbf u} = \begin{bmatrix} a^1 \ a^2 \dots \ a^m \end{bmatrix}\
\begin{bmatrix} {\mathbf e}_1 \\ {\mathbf e}_2\\ \vdots \\ {\mathbf e}_m\end{bmatrix}, \quad
{\mathbf w} = \begin{bmatrix} b^1 \ b^2 \dots \ b^n\end{bmatrix} \
\begin{bmatrix} \tilde{\mathbf e}_1 \\ \tilde{\mathbf e}_2\\ \vdots \\ \tilde{\mathbf e}_n \end{bmatrix}, \label{dsum1}
\end{equation}
where the coordinates $\left\{ a^i, b^j \in {\mathcal F} \mid i = 1, 2, \dots, m, \ j =1, 2, \dots, n \right\}$, then, 
\begin{equation}
{\mathbf v} =  {\mathbf u} \oplus {\mathbf w} = 
\begin{bmatrix} a^1 \ a^2 \dots \ a^m \ b^1 \ b^2 \dots \ b^n \end{bmatrix} 
\begin{bmatrix} {\mathbf e}_1 \\ {\mathbf e}_2\\ \vdots \\ {\mathbf e}_m\\
\tilde{\mathbf e}_1 \\ \tilde{\mathbf e}_2\\ \vdots \\ \tilde{\mathbf e}_n \end{bmatrix}.
\end{equation}

\subsubsection{Bilinear Maps and Bilinear Functionals}

For three vector spaces ${\bm V}$, ${\bm W}$, and ${\bm U}$ over ${\mathcal F}$, consider the function, 
\begin{eqnarray}
\phi : {\bm V}  \times {\bm W} & \to & {\bm U} \label{cartmap}\\
\left( {\mathbf v}, {\mathbf w} \right)& \mapsto & \phi \left( {\mathbf v}, {\mathbf w} \right),  
\end{eqnarray} 
where $\left({\mathbf v}, {\mathbf w} \right) $ is an ordered pair,
\begin{equation}
{\bm V} \times {\bm W} = \bigl\{ \left( {\mathbf v}, {\mathbf w} \right) \mid {\mathbf v} \in {\bm V}, \ {\mathbf w} \in {\bm W} \bigr\}.
\end{equation}
The function $\phi$ is a {\it bilinear map} or {\it bilinear transformation} if and only if, 
\begin{itemize}
\item[(i)] \ for all $\tilde{\mathbf v} \in {\bm V}$: \quad   ${\mathbf w}  \mapsto  \phi \left( {\tilde{\mathbf v}}, {\mathbf w} \right)$ is a {\it linear map}, and, 
\item[(ii)] \ for all $\tilde{\mathbf w} \in {\bm W}$: \quad  ${\mathbf v}  \mapsto  \phi \left( {\mathbf v}, {\tilde{\mathbf w}} \right)$ is a {\it linear map}. 
\end{itemize}
Namely, a bilinear map is a function that is a linear map in each of its two arguments. 
It is important to note that the bilinear map acts on a Cartesian product of two vector spaces.
There is no algebraic structure associated with a bilinear map. The Cartesian product of two
vector spaces with an algebraic structure follows the rules in Section \ref{sub2:linearmap}

In a manner similar to linear functionals, we define a {\it bilinear functional},
\begin{equation}
f_B : {\bm V} \times {\bm W} \to {\mathcal F} \Rightarrow \left( {\mathbf v}, {\mathbf w} \right) \mapsto  f_B \left( {\mathbf v}, {\mathbf w} \right), \quad 
{\mathbf v} \in {\bm V}, \ {\mathbf w} \in {\bm W},
\end{equation}
having the following properties,
\begin{itemize}
\item $f_B \left( {\mathbf v} + {\mathbf u}, {\mathbf w} \right) = f_B \left( {\mathbf v}, {\mathbf w} \right) + f_B \left( {\mathbf u}, {\mathbf w} \right)$, \ \ 
$\forall$ \ ${\mathbf u} \in {\bm V}$
\item $f_B \left( {\mathbf v}, {\mathbf w} + {\mathbf u} \right) = f_B \left( {\mathbf v}, {\mathbf w} \right) + f_B \left( {\mathbf v}, {\mathbf u} \right)$, \ \ 
$\forall$ \  ${\mathbf u} \in {\bm W}$
\item $f_B \left( {a \mathbf v}, {\mathbf w} \right) = a^* f_B \left( {\mathbf v}, {\mathbf w} \right)$
and $f_B \left( {\mathbf v}, b {\mathbf w} \right) = b f_B \left( {\mathbf v}, {\mathbf w} \right)$, $\forall \ a, b \in {\mathcal F}$, where
$a^*$ is the complex conjugate of $a$.
\end{itemize}

If ${\bm V}^*$ and ${\bm W}^*$ are the dual spaces corresponding to ${\bm V}$ and ${\bm W}$, respectively, the
bilinear functional on ${\bm V}^* \times {\bm W}^*$ is,
\begin{equation}
 {\bm V}^* \times {\bm W}^* \to {\mathcal F} \quad \Longrightarrow \quad \left( {\mathfrak f},  {\mathfrak g} \right) \mapsto 
{\mathfrak f} \left( {\mathbf v} \right) {\mathfrak g} \left( {\mathbf w} \right), \label{bilin_last}
\end{equation}
where ${\mathfrak f} \in {\bm V}^*$, 
${\mathfrak g} \in {\bm W}^*$, ${\mathbf v} \in {\bm V}$, and ${\mathbf w} \in {\bm W}$. We denote the vector space
of all bilinear functionals on ${\bm V} \times {\bm W}$ as ${\mathscr B} \left( {\bm V}, {\bm W}, {\mathcal F} \right)$; that is,
the bilinear functional on ${\bm V} \times {\bm W}$ is an element in ${\mathscr B} \left( {\bm V}, {\bm W}, {\mathcal F} \right)$.

\subsubsection{Tensor Product}
\label{subsub:tensorprod}

Consider an algebraic structure ${\bm V} \otimes {\bm W}$ that connects two vector spaces, ${\bm V}$ and ${\bm W}$ over $\mathcal F$, through
a binary operation $\otimes$ with the following properties,
\begin{itemize}
\item[(i)] \ \ $ {\mathbf v} \otimes \left( {\mathbf w}_1+ {\mathbf w}_2 \right) = {\mathbf v} \otimes {\mathbf w}_1 + {\mathbf v} \otimes {\mathbf w}_2 $
\item[(ii)] \ \  $ \left( {\mathbf v}_1 + {\mathbf v}_2 \right) \otimes {\mathbf w} = {\mathbf v}_1 \otimes {\mathbf w} + {\mathbf v}_2 \otimes {\mathbf w} $
\item[(iii)] \ \ \ $ a\left( {\mathbf v} \otimes  {\mathbf w} \right) = \left( a {\mathbf v} \right) \otimes  {\mathbf w}  =  {\mathbf v} \otimes  \left( a {\mathbf w} \right) $, 
\end{itemize}
for all $ {\mathbf v}, {\mathbf v}_1, {\mathbf v}_2  \in {\bm V}$, 
$ {\mathbf w}, {\mathbf w}_1, {\mathbf w}_2  \in {\bm W}$, and $a \in {\mathcal F}$. 
It follows that ${\bm V} \otimes {\bm W}$ is a vector space, referred to as a {\it tensor product} of two vector spaces. 

It is particularly useful to assign the tensor product ${\bm V} \otimes {\bm W}$ as an element in the vector space of bilinear functionals 
${\mathscr B} \left( {\bm V}, {\bm W}, {\mathcal F} \right)$
on ${\bm V} \times {\bm W}$. 
Consequently, for every ${\mathbf v} \in {\bm V}$ and ${\mathbf w} \in {\bm W}$, the tensor product ${\mathbf v} \otimes {\mathbf w} \in {\bm V} \otimes {\bm W}$ 
is such that,
\begin{equation}
{\mathbf v} \otimes {\mathbf w}: {\bm V}^* \times {\bm W}^* \to {\mathcal F} \quad \Longrightarrow \quad 
\left( {\mathfrak f}, {\mathfrak g} \right) \mapsto {\mathfrak f} \left( {\mathbf v} \right) {\mathfrak g} \left( {\mathbf w} \right),
\label{tenprop}
\end{equation}
for all ${\mathfrak f} \in {\bm V}^*$ and ${\mathfrak g} \in {\bm W}^*$. That this form is similar to \eqref{bilin_last} is not a coincidence; as we will
later show, maps of the Cartesian product are related to the maps of the tensor product.

For the vector spaces ${\bm V}$ and ${\bm W}$ leading to \eqref{dsum1}, the tensor product can be written in 
a compact form, 
\begin{equation}
{\mathbf v} \otimes {\mathbf w} =\sum_{i=1}^m \ \sum_{j=1}^n \, a^i b^j  \left( {\mathbf e}_i \otimes \tilde{\mathbf e}_j \right).\label{tp1}
\end{equation}
The set $ \left\{ {\mathbf e}_i \otimes \tilde{\mathbf e}_j \ \mid \ i = 1, \dots, m; \ j = 1, \dots, n \right\} $  is the basis set of, and
spans, the vector space $ {\bm V} \otimes {\bm W} $;  $ {\rm dim} \left( {\bm V} \otimes {\bm W} \right) = 
{\rm dim} \left( {\bm V} \right) {\rm dim}  \left( {\bm W} \right) $. The association of this product with a tensor is 
due to the elements in the basis set having two indices.

Any element ${\mathbf u}$ in the vector space ${\bm V} \otimes {\bm W}$ will have a unique representation,
\begin{equation}
{\mathbf u} = \sum_{i=1}^m \ \sum_{j=1}^n \ c^{ij}  \left( {\mathbf e}_i \otimes \tilde{\mathbf e}_j \right), \label{tp2}
\end{equation}
where the coordinates $c^{ij} \in {\mathcal F}$. In general, $c^{ij}$ will not split into the  product $a^i b^j$ in \eqref{tp1}. Consequently, 
the tensor product is not surjective (onto). The following simple example illustrates this point.

If ${\rm dim}({\bm V}) = {\rm dim}({\bm W}) = 2$,  \eqref{tp1} has the general form,
\begin{equation}
{\mathbf v} \otimes {\mathbf w} = a^1 b^1  \left( {\mathbf e}_1 \otimes \tilde{\mathbf e}_1 \right)  + 
a^1 b^2  \left( {\mathbf e}_1 \otimes \tilde{\mathbf e}_2 \right) + 
a^2 b^1  \left( {\mathbf e}_2 \otimes \tilde{\mathbf e}_1 \right) + 
a^2 b^2  \left( {\mathbf e}_2 \otimes \tilde{\mathbf e}_2 \right),  \label{tp3}
\end{equation} 
while \eqref{tp2} becomes,
\begin{equation}
{\mathbf u} = c^{11}  \left( {\mathbf e}_1 \otimes \tilde{\mathbf e}_1 \right)  +
c^{12}  \left( {\mathbf e}_1 \otimes \tilde{\mathbf e}_2 \right) +
c^{21}  \left( {\mathbf e}_2 \otimes \tilde{\mathbf e}_1 \right) +
c^{22}  \left( {\mathbf e}_2 \otimes \tilde{\mathbf e}_2 \right). \label{tp4}
\end{equation}
Since the elements in the basis set are linearly independent, equating \eqref{tp3} and \eqref{tp4} leads to four equations for the coordinates,
\begin{equation}
c^{11} = a^1 b^1, \quad c^{12} = a^1 b^2, \quad c^{21} = a^2 b^1, \quad c^{22} = a^2 b^2. \label{tp5}
\end{equation}
Let us consider an element ${\mathbf u}$ with $c^{11} = 0$ and $c^{12}, c^{21}, c^{22} \ne 0$.
We cannot express ${\mathbf u}$ as a tensor product ${\mathbf v} \otimes {\mathbf w}$ since $c^{11} = a^1 b^1 = 0$ implies
either $a^1 = 0$ or $b^1 = 0$. If $a^1 = 0$, the element ${\mathbf e}_1 \otimes \tilde{\mathbf e}_2$ in \eqref{tp3} does not exist. If
$b^1 = 0$, the element ${\mathbf e}_2 \otimes \tilde{\mathbf e}_1$ is not present. Hence, the chosen ${\mathbf u}$  cannot be represented
by any tensor product ${\mathbf v} \otimes {\mathbf w}$.

\paragraph{\bf Basis for the Tensor Product of Two Vector Spaces}

For the two vector spaces ${\bm V}$ and ${\bm W}$ with
${\rm dim} \left( {\bm V} \right) = n$ and ${\rm dim} \left( {\bm W} \right) = m$, assume their
basis sets are $\left\{ {\mathbf e}_{vi} \mid  i = 1, 2, \dots, n \right\}$ and
$\left\{ {\mathbf e}_{wi} \mid i = 1, 2, \dots, m \right\}$, respectively. Let $\left\{ \left.  {\mathbf f}_{v}^i \ \right| i = 1, 2, \dots, n \right\}$ and
$\left\{ {\mathbf f}_{w}^i \mid i = 1, 2, \dots, m \right\}$ be the basis sets of the dual vector spaces ${\bm V}^*$ and ${\bm W}^*$, respectively.
If $\left( {\mathfrak f}_v, {\mathfrak f}_w \right) \in {\bm V}^* \times {\bm W}^*$ and ${\mathbf T} \in {\bm V} \otimes {\bm W}$, then,
\begin{equation}
{\mathbf T} \left( {\mathfrak f}_v, {\mathfrak f}_w \right)  =  {\mathbf T} \left( \sum\limits_{i=1}^n  \alpha_{vi} {\mathbf f}_v^i, 
\sum\limits_{j=1}^m  \alpha_{wj} {\mathbf f}_w^j \right) 
 =  \sum\limits_{i=1}^n \sum\limits_{j=1}^m \ \alpha_{vi} \alpha_{wj} {\mathbf T} \left( {\mathbf f}_v^i, {\mathbf f}_w^j \right), \label{ex1}
\end{equation}
where $\alpha_{vi}$ and $\alpha_{wj}$ are the respective coordinates, and we have made use of the
bilinearity of ${\mathbf T}$. 

Defining $T^{ij} = {\mathbf T} \left( {\mathbf f}_v^i, {\mathbf f}_w^j \right)$, we find that,
\begin{equation}
\sum\limits_{i=1}^n \sum\limits_{j=1}^m \ T^{ij} {\mathbf e}_{vi} \otimes {\mathbf e}_{wj} \left(  {\mathbf f}_v^k, {\mathbf f}_w^l \right) 
= \sum\limits_{i=1}^n \sum\limits_{j=1}^m \ T^{ij} {\mathbf f}_v^k \left( {\mathbf e}_{vi} \right) {\mathbf f}_w^l \left( {\mathbf e}_{wj} \right)
= \sum\limits_{i=1}^n \sum\limits_{j=1}^m \ T^{ij} \delta_i^k \delta_j^l = T^{kl} . \label{ex2}
\end{equation}
The first equality follows from \eqref{tenprop}, while the second equality is a property of dual spaces. 
Replacing $T^{ij}$ on the right hand side of \eqref{ex1} by the expression on the left hand side of \eqref{ex2}, and realizing that the
resulting equation is valid for any ${\mathfrak f}_v$ and ${\mathfrak f}_w$, we obtain,
\begin{equation}
{\mathbf T} = \sum\limits_{i=1}^n \ \sum\limits_{j=1}^m \ T^{ij} {\mathbf e}_{vi} \otimes {\mathbf e}_{wj}. \label{ex3}
\end{equation}
Thus, $\left\{ {\mathbf e}_{vi} \otimes {\mathbf e}_{wj} \mid i = 1, \dots n, \ j = 1, \dots m \right\}$ forms
the basis set of dimension $nm$ for the vector space ${\bm V} \otimes {\bm W}$.

\paragraph{\bf The Universal Property}

The map defined by, 
\begin{eqnarray}
\phi_{\otimes} : {\bm V}  \times {\bm W} & \to & {\bm V} \otimes {\bm W} \label{tpr1a}\\
\left( {\mathbf v}, {\mathbf w} \right) & \mapsto &{\mathbf v} \otimes {\mathbf w}, \label{tpr1b}
\end{eqnarray}
is a bilinear map taking into account properties of the tensor product $\otimes$. It turns out
that, given the map $\phi$ in \eqref{cartmap}, there exists a unique linear map,
\begin{equation}
f_\otimes :  {\bm V} \otimes {\bm W} \to {\bm U}, \label{tpbmap}
\end{equation}
 such that $\phi = f_\otimes \circ \phi_{\otimes}$.\footnote{$\circ$ indicates function composition of maps.}
This connects the mapping of the Cartesian product \eqref{cartmap} to the tensor product \eqref{tpr1a}, and is
referred to as the {\it universal property} of tensor products.

We can reformulate the universal property to convert a bilinear map to a linear map. For the bilinear map
$\phi$, \eqref{cartmap}, there exists a unique linear map $f_\otimes$, \eqref{tpbmap},  such that,
\begin{equation}
f_\otimes \left( {\mathbf v} \otimes {\mathbf w} \right) = \phi \left( {\mathbf v}, {\mathbf w} \right),
\end{equation}
for all ${\mathbf v} \in {\bm V}$ and ${\mathbf w} \in {\bm W}$. 
Conversely, for the linear map $f_{\otimes}$, \eqref{tpbmap},  there exists a unique bilinear map $\phi$, \eqref{cartmap}, such that,
\begin{equation}
\phi \left( {\mathbf v}, {\mathbf w} \right) = f_\otimes \left( {\mathbf v} \otimes {\mathbf w} \right),
\end{equation}
for all ${\mathbf v} \in {\bm V}$ and ${\mathbf w} \in {\bm W}$.

\subsection{\textbf{\textit{Multilinear Maps and Tensors}}}
\label{subs:multilinear}

The description of bilinear maps and functionals can be generalized to multilinear maps and functionals. For
vector spaces ${\bm W}$, $\left\{ {\bm V}_1, {\bm V}_2, \dots , {\bm V}_r \right\} $,\footnote{The subscripts indicate
different vector spaces.} $ r \in {\mathbb N} $, over the field ${\mathcal F}$, the function,
\begin{eqnarray}
{\phi}_M: {\bm V}_1 \times {\bm V}_2 \times \dots \times {\bm V}_r & \to & {\bm W} \\
\left( {\mathbf v}_1, {\mathbf v}_2, \dots , {\mathbf v}_r \right) & \mapsto & {\phi}_M \left( {\mathbf v}_1, {\mathbf v}_2, \dots , {\mathbf v}_r \right),  
\end{eqnarray} 
is a {\it multilinear map} if it is linear in each one of its variables while keeping the other variables unchanged; that is,
\begin{eqnarray}
 {\phi_M} \left( {\mathbf v}_1, {\mathbf v}_2, \dots , {\mathbf v}_{i-1}, a {\mathbf v}_i + b {\mathbf u}_i, {\mathbf v}_{i+1}, \dots , {\mathbf v}_r \right) = \nonumber \\
 a\, {\phi_M} \left( {\mathbf v}_1, {\mathbf v}_2, \dots , {\mathbf v}_{i-1},  {\mathbf v}_i , {\mathbf v}_{i+1}, \dots , {\mathbf v}_r \right) + \nonumber \\
 b\, {\phi_M} \left( {\mathbf v}_1, {\mathbf v}_2, \dots , {\mathbf v}_{i-1},  {\mathbf u}_i , {\mathbf v}_{i+1}, \dots , {\mathbf v}_r \right),
\end{eqnarray}
where ${\mathbf u}_i \in {\bm V}_i$, $i \in {\mathbb N}$, $1 \le i \le m$, and $a, \, b  \in {\mathcal F}$.

Analogously, we can define a {\it multilinear functional} or {\it multilinear form},
\begin{eqnarray}
{f_M}: {\bm V}_1 \times {\bm V}_2 \times \dots \times {\bm V}_r & \to & {\mathcal F} \\
\left( {\mathbf v}_1, {\mathbf v}_2, \dots , {\mathbf v}_r \right) & \mapsto &{f_M} \left( {\mathbf v}_1, {\mathbf v}_2, \dots , {\mathbf v}_r \right). 
\end{eqnarray}

We will usually be operating within a single vector space ${\bm V}$ and its dual space ${\bm V}^*$. Then,
\begin{equation}
{\mathbf T}: \underbrace{ {\bm V} \times {\bm V} \times \dots \times {\bm V}}_\text{r times}
\times \underbrace{ {\bm V}^* \times {\bm V}^* \times \dots \times {\bm V}^* }_\text{s times} \to  {\mathcal F} \label{ten1a}
\end{equation} 
is a {\it tensor of order} $\left( r, s \right)$ on ${\bm V}$; \ $r,  s  \in {\mathbb N}$. 
The set of all tensors ${\mathbf T}$ is denoted by ${\mathcal T}^r_s( {\bm V})$. The {\it rank} of the tensor  
is $r+s$, and is independent of the dimension of the underlying vector space ${\bm V}$. 
 
A few remarks about tensors for some special cases, 
\begin{enumerate}
\item[(1)] $r=0, \ s=0$: by convention, ${T}: {\mathcal F} \to {\mathcal F}$, which implies that a tensor of rank 0, 
${\mathcal T}^0_0$, is a scalar --  a set of all elements $\in {\mathcal F}$. 
\item[(2)] $r = 1, \ s = 0$: from \eqref{mapfunc}, the map ${\mathbf T}: {\bm V} \to {\mathcal F}$ defines a linear functional  
on ${\bm V}$ -- the dual space. Thus, ${\mathcal T}^1 ( {\bm V} ) = {\bm V}^*$ is a covector -- a covariant vector. 
\item[(3)] $r = 0, \ s = 1$: the map ${\mathbf T}: {\bm V}^* \to {\mathcal F}$ defines a linear functional  
on ${\bm V}^*$, denoted by ${\bm V}^{**}$.\footnote{The dual of a vector space is a vector space in its own right. Consequently,  
it is reasonable to consider the dual of a dual space.} Thus, ${\mathcal T}_1 ( {\bm V} ) = {\bm V}^{**} \cong {\bm V}$ 
is a vector -- a contravariant vector.\footnote{A finite dimensional vector space ${\bm V}$ is isomorphic to its double dual vector space ${\bm V}^{**}$.}
\end{enumerate}
From the above remarks, it follows that
$r$ is referred to as the covariant index and $s$ as the contravariant index. 
For $r = 0$,  the contravariant tensor is written as ${\mathcal T}_s ({\bm V})$, while for $s=0$, the covariant tensor is
${\mathcal T}^r ({\bm V})$. For $r, \ s \ne 0$, ${\mathcal T}^r_s ({\bm V})$ is referred to as a mixed tensor. 
 
\paragraph{\bf Basis Set for a Tensor}

The discussion in Section \ref{subsub:tensorprod} can be readily extended to multilinear maps. Since the tensor product is  
an element in the vector space of multilinear functionals on the dual vector space of the Cartesian product, \eqref{ten1a} can  
be written as a tensor product, 
\begin{equation}
{\mathbf T}: \underbrace{ {\bm V}^* \otimes {\bm V}^* \otimes \dots \otimes {\bm V}^*}_\text{r times}
\otimes \underbrace{ {\bm V} \otimes {\bm V} \otimes \dots \otimes {\bm V} }_\text{s times} \to  {\mathcal F}. \label{ten1b}
\end{equation} 

Following steps analogous to those in the latter part of Section \ref{subsub:tensorprod}, any tensor ${\mathbf T} \in {\mathcal T}^r_s$ can be
written as,
\begin{equation}
{\mathbf T} = \sum\limits_{i_1=1}^n \dots \sum\limits_{i_r=1}^n   \ \sum\limits_{j_1=1}^n \dots \sum\limits_{j_s=1}^n \ T_{i_1 \dots i_r}^{j_1 \dots j_s} \
{\mathbf f}^{i_1} \otimes \dots \otimes {\mathbf f}^{i_r} \otimes {\mathbf e}_{j_1} \otimes \dots 
\otimes {\mathbf e}_{j_s}, \label{tenful1}
\end{equation}
where,
\begin{equation}
\left\{ 
{\mathbf f}^{i_1} \otimes \dots \otimes {\mathbf f}^{i_r} \otimes {\mathbf e}_{j_1} \otimes \dots 
\otimes {\mathbf e}_{j_s}
\mid i_1, \dots i_r, j_1 \dots j_s  = 1, \dots , n \right\}, \label{tenful2}
\end{equation}
is a basis set of dimension $n^{r+s}$, with ${\rm dim} \left( {\bm V} \right) = n$.
The identification of each element in the basis set with the corresponding ${\bm V}^*$ and ${\bm V}$ in \eqref{ten1b}
is quite clear. Note that,
\begin{equation}
{\mathbf T} \left( {\mathbf e}_{i_1}, \dots , {\mathbf e}_{i_r}, {\mathbf f}^{j_1}, \dots , {\mathbf f}^{j_s} \right) = T_{i_1 \dots i_r}^{j_1 \dots j_s}.
\label{tenful3}
\end{equation}
At this stage, the analogy between covariant/contravariant vectors and tensors becomes apparent. For a covariant vector the corresponding
coordinates were represented by subscripts. In \eqref{tenful1}, for $s=0$, the ``coordinates'' $T_{i_1 \dots i_r}$ are represented by
subscripts and the associated tensor ${\mathcal T}^r$ is covariant. For contravariant vectors and tensors the coordinates are represented by superscripts,
and ${\mathcal T}_s$ is a contravariant tensor.
 
\paragraph{\bf Rank 2 Mixed Tensor as an Operator}

Consider a tensor ${\mathbf T} \in {\mathcal T}_1^1 ( {\bm V} )$ with ${\mathbf T}: {\bm V}^* \otimes {\bm V}\to {\mathcal F}$.
Then $\left( {\mathfrak f}, {\mathbf v} \right) \in {\bm V}^* \otimes {\bm V}$. 
Following the discussion in Section \ref{subsub:tensorprod}, given that,
\begin{equation}
\sum_{i, j = 1}^n T_i^j {\mathbf f}^i \otimes {\mathbf e}_j \in {\bm V}^* \otimes {\bm V},
\end{equation}
we define a linear operator $T$ by,
\begin{equation}
T \left( {\mathbf v} \right) = \sum_{i, j = 1}^n T_i^j {\mathbf f}^i \left( {\mathbf v} \right)  {\mathbf e}_j 
= \sum_{i, j = 1}^n T_i^j v^i {\mathbf e}_j,
\end{equation}
where ${\mathbf v} = v^1 {\mathbf e}_1 + \dots + v^n  {\mathbf e}_n \ \in \ {\bm V}$.
This form of $T$ is exactly the same as that of the linear operator \eqref{linop3} in Section \ref{sub2:linearmap}.

\subsubsection{Metric Tensor:  Transforming Between Vectors and Covectors}
\label{sub:metric}

In physics, at times it is convenient to ``transform'' tensors in a way that one or more of the indices 
are either raised or lowered, for example, replacing $T^{ij}$ by $\bar{T}^{i}_j$. From \eqref{ten1a}, \eqref{ten1b}, \eqref{tenful1}, and
\eqref{tenful3}, it is evident that lowering or raising a tensor index requires replacing of a vector space by its dual vector space
or vice-versa.  

Consider a bilinear functional ${\mathit g}: {\bm V} \times {\bm V} \to {\mathcal F}$; that is, ${\mathit g}$ is an element in
the bilinear dual space.  Note that ${\mathit g}$ is also a rank 2 
covariant tensor ${\mathit g} \in {\mathcal T}^2 \left( {\bm V} \right)$; ${\mathit g}: {\bm V}^* \otimes {\bm V}^* \to {\mathcal F}$.
Following the notation in Section \ref{subs:multilinear},
\begin{equation}
{\mathit g} = \sum\limits_{i, j = 1}^{n} g_{ij} {\mathbf f}^i \otimes {\mathbf f}^j, \label{met1}
\end{equation}
where $g_{ij} = {\mathit g} \left( {\mathbf e}_i, {\mathbf e}_j \right)$. Since, ${\mathit g}$ is a bilinear map from identical vector
spaces to ${\mathcal F}$, ${\mathit g}$ is a symmetric bilinear form and $g_{ij} = g_{ji}$.

For any ${\mathbf v} \in {\bm V}$,
\begin{align}
{\mathit g} \left( {\mathbf v} \right)  = & \sum\limits_{i, j = 1}^{n} g_{ij} {\mathbf f}^i \otimes {\mathbf f}^j 
\underbrace{ \left( \sum\limits_{k = 1}^n v^k {\mathbf e}_k \right)}_{=\ {\mathbf v}} = \sum\limits_{i, j, k = 1}^{n} g_{ij}  v^k {\mathbf f}^i \delta^j_k \nonumber \\
 = & \sum\limits_{i = 1}^{n}\ \underbrace{ \left( \sum\limits_{j = 1}^{n} g_{ij} v^j \right) }_{{\rm define \ as \ } v_i^\prime}  {\mathbf f}^i = 
\underbrace{ \sum\limits_{i = 1}^{n} v_i^\prime {\mathbf f}^i }_{{\rm define \ as \ } {\mathbf v}^\prime} \ \in {\bm V}^*
\label{met2}
\end{align}
The tensor ${\mathit g}$ is referred to as the {\it metric tensor} and its action transforms
a vector into a covector in the dual space.  By operating on a tensor \eqref{ten1b} with
${\mathit g}$, we convert one ${\bm V}$ to ${\bm V}^*$, and
the set of all tensors changes from ${\mathcal T}^r_s$ to ${\mathcal T}^{r+1}_{s-1}$. Then the right hand side of \eqref{tenful1} becomes,
\begin{equation}
\sum\limits_{i_1=1}^n \dots \sum\limits_{i_{r+1}=1}^n   \ \sum\limits_{j_1=1}^n \dots \sum\limits_{j_{s-1}=1}^n \ T_{i_1 \dots i_{r+1}}^{j_1 \dots j_{s-1}} \
{\mathbf f}^{i_1} \otimes \dots \otimes {\mathbf f}^{i_{r+1}} \otimes {\mathbf e}_{j_1} \otimes \dots 
\otimes {\mathbf e}_{j_{s-1}}. \label{met3}
\end{equation}
We notice that one of the indices of $T$ in \eqref{met3} has been lowered.
Each application of $g$ transforms a vector space to its dual space thereby lowering another index.

Similarly, we can show that a metric tensor $\bar{\mathit g}: {\bm V} \otimes {\bm V} \to {\mathcal F}$,
\begin{equation}
\bar{\mathit g}= \sum\limits_{i, j = 1}^{n} \bar{g} ^{ij} {\mathbf e}_i \otimes {\mathbf e}_j, \label{met4}
\end{equation}
transforms a covector to its vector, thereby, raising the index of a tensor.
Here $\bar{g}^{ij} = \bar{g} \left( {\mathbf f}^i, {\mathbf f}^j \right)$. For ${\mathbf w}' \in {\bm V}^*$,
\begin{equation}
\bar{\mathit g} \left( {\mathbf w}' \right) = 
\sum\limits_{i, j = 1}^{n} \bar{g}^{ij} {\mathbf e}_i \otimes {\mathbf e}_j 
\underbrace{ \left( \sum\limits_{k = 1}^n w'_k {\mathbf f}^k \right)}_{=\ {\mathbf w}'} = 
\sum\limits_{i = 1}^{n}\ \underbrace{ \left( \sum\limits_{j = 1}^{n} \bar{g}^{ij} w'_j \right) }_{{\rm define \ as \ } w^i}  {\mathbf e}_i = 
\underbrace{ \sum\limits_{i = 1}^{n} w^i {\mathbf e}_i }_{{\rm define \ as \ } {\mathbf w}} \ \in {\bm V}. \label{met5}
\end{equation}
The consistency condition $\bar{\mathit g} \left( {\mathit g} \left( {\mathbf v} \right) \right) = {\mathbf v} $ implies
that $\bar{{\mathit g}}^{-1} = {\mathit g}$. For the inverse to exist, it requires that the map ${\mathit g}: {\bm V} \times {\bm V} \to {\mathcal F}$
be non-degenerate; that is, for ${\mathbf v}, {\mathbf w} \in {\bm V}$, $g \left( {\mathbf v}, {\mathbf w} \right) = 0$ 
for all ${\mathbf v}$ implies ${\mathbf w} = 0$.

The rank 2 tensor ${\mathit g}$ is covariant with its elements represented by ${\mathit g}_{ij}$, while $\bar{\mathit g}$ is a contravariant tensor
with elements given by $\bar{\mathit g}^{ij}$. Note that both tensors are symmetric. Furthermore, since $\bar{{\mathit g}}^{-1} = {\mathit g}$,
\begin{equation}
\sum\limits_{j =1}^n \ {\mathit g}_{i j} \bar{\mathit g}^{j k} = \delta_i^k. \label{met6}
\end{equation}

\subsection{\textbf{\textit{Transforming Between Bases of Vector Spaces: Tensors and Matrix Representations}}}
\label{sub:transform}

In classical electromagnetism and in plasma physics, we usually encounter tensors as matrices -- for example,
the Maxwell stress tensor for electromagnetic waves or the linear permittivity tensor for a magnetized plasma.
However, tensors are more diverse and play an important role in physics. 
While we are free to choose the basis of a vector space that is
convenient for a given situation, the laws of physics are independent of the basis functions. Furthermore, tensors
are independent of reference frames. In transforming classical
Maxwell equations to a form suitable for quantum computers, we need to make sure that the underlying physics, such
as that associated with conservation laws, is preserved. Tensor algebra provides a convenient conduit for a wide range
of vector space transformations while preserving the underlying physics. 
A tensor space is a generalization of a vector space with additional transformation properties. As we have shown,
the roots of tensors are in multilinear algebra -- linear algebra with multiple vector spaces or multiple replicates of the same
vector space operating in unison. 

For a vector space ${\bm V}$ with ${\rm dim}({\bm V}) = n$, let $\left\{ {\mathbf e}_1, {\mathbf e}_2, \dots , {\mathbf e}_n\right\}$ and
$\left\{ \tilde{\mathbf e}_1, \tilde{\mathbf e}_2, \dots , \tilde{\mathbf e}_n \right\}$ be two different sets of basis functions. 
In each set, a vector ${\mathbf v} \in {\bm V}$ takes on the form,
\begin{eqnarray}
{\mathbf v} & =  \sum\limits_{i=1}^n \ v^i {\mathbf e}_i & =  \begin{bmatrix} v^1 &  v^2 & \dots &  v^n \end{bmatrix} 
\begin{bmatrix}  {\mathbf e}_1 \\ {\mathbf e}_2 \\ \vdots \\ {\mathbf e}_n \end{bmatrix}, \label{tt1} \\
{\mathbf v} & =  \sum\limits_{j=1}^n \ \tilde{v}^j \tilde{\mathbf e}_j & =  \begin{bmatrix} \tilde{v}^1 & \tilde{v}^2 & \dots & \tilde{v}^n \end{bmatrix} 
\begin{bmatrix}  \tilde{\mathbf e}_1 \\ \tilde{\mathbf e}_2 \\ \vdots \\ \tilde{\mathbf e}_n \end{bmatrix}, \label{tt2}
\end{eqnarray}
where $\left\{ v^1, v^2, \dots, \ v^n \right\} \in {\mathcal F}$ and 
$\left\{ \tilde{v}^1, \tilde{v}^2, \dots, \ \tilde{v}^n \right\} \in {\mathcal F}$ are sets of coordinates of ${\mathbf v}$ 
in the two sets of basis functions, respectively. 
Furthermore, based on  the properties of basis functions, we can uniquely express one set of basis functions in terms of the other set,
\begin{equation}
\tilde{\mathbf e}_j = \sum\limits_{i=1}^n \ {\mathrm R}^i_j {\mathbf e}_i, \label{tt3a}
\end{equation}
where ${\mathrm R}^i_j$ are elements of a mixed tensor ${\mathcal R}$ of rank 2; ${\mathcal R} = \left[ {\mathrm R}^i_j \right] $ with
${\mathcal R}: {\bm V} \to {\bm V}$. Equivalently,  ${\mathbf v} \mapsto {\mathcal R} \left( {\mathbf v} \right)$ with ${\mathbf v} \in {\bm V}$. 
We can write \eqref{tt3a} in a matrix form,\footnote{A rank 2 tensor can
be written as a matrix but not all matrices are rank 2 tensors.} 
\begin{equation}
\begin{bmatrix}  \tilde{\mathbf e}_1 \\[6pt] \tilde{\mathbf e}_2 \\ \vdots \\ \tilde{\mathbf e}_n \end{bmatrix} \ = \
\begin{bmatrix} {\mathrm R}^1_1 & {\mathrm R}^2_1 & \dots & {\mathrm R}^n_1 \\[6pt] 
{\mathrm R}^1_2 & {\mathrm R}^2_2 & \dots & {\mathrm R}^n_2 \\ 
\vdots & \vdots & \ddots & \vdots \\ 
{\mathrm R}^1_n & {\mathrm R}^2_n & \dots & {\mathrm R}^n_n 
\end{bmatrix}
\begin{bmatrix}  {\mathbf e}_1 \\[6pt] {\mathbf e}_2 \\ \vdots \\ {\mathbf e}_n \end{bmatrix}. \label{tt3}
\end{equation} 
The transformation of the coordinates is given by ${\mathcal R}^{-1}$ -- the inverse of ${\mathcal R}$,
\begin{equation}
\begin{bmatrix}  \tilde{v}^1 & \tilde{v}^2 & \dots & \tilde{v}^n \end{bmatrix} \ = \
\begin{bmatrix}  {v}^1 & {v}^2 & \dots & {v}^n \end{bmatrix}
\begin{bmatrix} {\mathrm R}^1_1 & {\mathrm R}^2_1 & \dots & {\mathrm R}^n_1 \\[6pt] 
{\mathrm R}^1_2 & {\mathrm R}^2_2 & \dots & {\mathrm R}^n_2 \\ 
\vdots & \vdots & \ddots & \vdots \\ 
{\mathrm R}^1_n & {\mathrm R}^2_n & \dots & {\mathrm R}^n_n 
\end{bmatrix}^{-1}. \label{tt4}
\end{equation} 
Note that \eqref{tt3} and \eqref{tt4} lead to $\sum\limits_{i=1}^n \tilde{v}^i \tilde{\mathbf e}_i = \sum\limits_{i=1}^n {v}^i {\mathbf e}_i$, 
as expected.

The transformation of coordinates can be understood within the realm of the dual space ${\bm V}^*$. 
Let $\left\{ {\mathbf f}^1, {\mathbf f}^2, \dots, {\mathbf f}^n \right\}$ and 
$\left\{ \tilde{\mathbf f}^1, \tilde{\mathbf f}^2, \dots, \tilde{\mathbf f}^n \right\}$ be two basis sets in ${\bm V}^*$,
which correspond to the basis sets $\left\{ {\mathbf e}_1, {\mathbf e}_2, \dots , {\mathbf e}_n\right\}$ and
$\left\{ \tilde{\mathbf e}_1, \tilde{\mathbf e}_2, \dots , \tilde{\mathbf e}_n \right\}$ in ${\bm V}$, respectively. From \eqref{dual3},
\begin{equation}
{\mathbf f}^i \left( {\mathbf e}_j \right) = \delta^i _j, \quad \tilde{\mathbf f}^i \left( \tilde{\mathbf e}_j \right) = \delta^i _j.
\end{equation}
From \eqref{tt1} and \eqref{tt2},
\begin{equation}
{\mathbf f}^i \left( {\mathbf v} \right) = v^i, \quad \tilde{\mathbf f}^j \left( {\mathbf v} \right) = \tilde{v}^j, \label{tt5}
\end{equation}
where we have made use of \eqref{dual4}. We can express one set of basis functions in ${\bm V}^*$ in terms of the other set,
\begin{equation}
\tilde{\mathbf f}^i = \sum_{j=1}^n  \overline{{\mathrm R}}^i_j  {\mathbf f}^j \label{tt6},
\end{equation}
where $\overline{{\mathrm R}}^i_j$ are elements of a mixed tensor $\overline{\mathcal R}$ of rank 2; 
$\overline{\mathcal R} = \left[ \overline{{\mathrm R}}^i_j \right] $ with
$\overline{\mathcal R}: {\bm V}^* \to {\bm V}^*$.
Then,
\begin{equation}
\tilde{\mathbf f}^i \left( {\mathbf v} \right) = \tilde{v}^i = \sum_{j=1}^n  \overline{{\mathrm R}}^i_j  {\mathbf f}^j \left( {\mathbf v} \right) = 
\sum_{j=1}^n  \overline{{\mathrm R}}^i_j  v^j,
\label{tt7}
\end{equation}
where we have made use of \eqref{tt5}. It follows that,
\begin{equation}
\begin{bmatrix}  \tilde{v}^1 & \tilde{v}^2 & \dots & \tilde{v}^n \end{bmatrix} \ = \
\begin{bmatrix}  {v}^1 & {v}^2 & \dots & {v}^n \end{bmatrix}
\begin{bmatrix} \overline{{\mathrm R}}^1_1 & \overline{{\mathrm R}}^2_1 & \dots & \overline{{\mathrm R}}^n_1 \\[6pt]
\overline{{\mathrm R}}^1_2 & \overline{{\mathrm R}}^2_2 & \dots & \overline{{\mathrm R}}^n_2 \\ 
\vdots & \vdots & \ddots & \vdots \\
\overline{{\mathrm R}}^1_n & \overline{{\mathrm R}}^2_n & \dots & \overline{{\mathrm R}}^n_n 
\end{bmatrix}. \label{tt8}
\end{equation}
For the results in \eqref{tt8} and \eqref{tt4} to be consistent, we require that ${\mathcal R}^{-1} = \overline{\mathcal R}$.
Thus, the transformation of basis vectors is in vector space while the transformation of coordinates is in the dual space.

There is a slightly different way to derive the relation between ${\mathcal R}$ and $\overline{\mathcal R}$. Given the transformation
equations,
\begin{equation}
\tilde{\mathbf e}_i = \sum\limits_{k=1}^n {\mathrm R}^k_i {\mathbf e}_k, \quad \tilde{\mathbf f}^j = \sum\limits_{l=1}^n \overline{\mathrm R}^j_l {\mathbf f}^l,
\end{equation}
we obtain,
\begin{equation}
\tilde{\mathbf f}^j \left( \tilde{\mathbf e}_i \right)= \sum\limits_{l=1}^n \sum\limits_{k=1}^n \overline{\mathrm R}^j_l {\mathrm R}^k_i {\mathbf f}^l 
\left( {\mathbf e}_k \right).
\end{equation}
If we want to preserve the orthonormality relations, $\tilde{\mathbf f}^j \left( \tilde{\mathbf e}_i \right) = \delta^j_i$ and
${\mathbf f}^l \left( {\mathbf e}_k \right) = \delta^l_k$, then,
\begin{equation}
\sum\limits_{l=1}^n {\mathrm R}^l_i \overline{\mathrm R}^j_l= \delta^j_i \quad \Rightarrow \quad {\mathcal R} \overline{\mathcal R} = {\mathcal I}_n.
\label{transiden}
\end{equation}
Thus, ${\mathcal R}^{-1} = \overline{\mathcal R}$.

\subsubsection{Transformation of a Rank 2 Contravariant Tensor}
\label{sub:ex1}

Consider a rank 2 contravariant tensor ${\mathbf T} \in {\mathcal T}_2 \left( {\bm V} \right)$ with ${\mathbf T}: {\bm V}^* \times {\bm V}^* \to {\mathcal F}$. 
From \eqref{tenful1} and \eqref{tenful2}, 
\begin{equation}
{\mathbf T} = \sum_{i, j=1}^n \ {\bf T} \left( \tilde{\mathbf f}^i, \tilde{\mathbf f}^j \right) 
\tilde{\mathbf e}_i \otimes \tilde{\mathbf e}_j, \label{rank21}
\end{equation}
where we have chosen $\left\{ \tilde{\mathbf e}_1, \dots , \tilde{\mathbf e}_n \right\}$ and $\left\{ \tilde{\mathbf f}^1, \dots , \tilde{\mathbf f}^n \right\}$
as the basis sets for ${\bm V}$ and ${\bm V}^*$, respectively. If we transform to the basis sets
$\left\{ {\mathbf e}_1, \dots , {\mathbf e}_n \right\}$ and $\left\{ {\mathbf f}^1, \dots , {\mathbf f}^n \right\}$, respectively, using
\eqref{tt3a} and \eqref{tt6}, then \eqref{rank21} leads to,
\begin{equation}
{\mathbf T} = \sum_{i, j=1}^n  \left(  \sum_{k', l'=1}^n  \overline{\mathrm R}_{k'}^i 
\overline{\mathrm R}_{l'}^j {\mathbf T} \left( {\mathbf f}^{k'}, {\mathbf f}^{l'} \right) \right)
\left(  \sum_{k, l=1}^n \ {\mathrm R}_i^k 
{\mathrm R}_j^l  {\mathbf e}_{k} \otimes {\mathbf e}_l \right). \label{rank22}
\end{equation}
From \eqref{transiden},
\begin{equation}
\sum\limits_{i=1}^n \overline{\mathrm R}_{k'}^i {\mathrm R}_i^k = \delta_{k'}^k, \quad 
\sum\limits_{j=1}^n \overline{\mathrm R}_{l'}^j {\mathrm R}_j^l = \delta_{l'}^l, \label{rank23}
\end{equation}
so that \eqref{rank22} reduces to,
\begin{equation}
{\mathbf T} = \sum_{k,l=1}^n \  {\bf T} \left( {\mathbf f}^k, {\mathbf f}^l \right) 
{\mathbf e}_k \otimes {\mathbf e}_l \label{rank24}.
\end{equation}
Upon comparing with \eqref{rank21}, we note that tensors remain invariant under transformations between
two different basis sets and their duals. Consequently, it is useful to express physical laws in terms of tensors
as we are not constrained to a particular choice of basis sets.

\subsection{\textbf{\textit{Inner Product Space, Metric Space, Hilbert Space}}}
\label{sub:innpro}

Let us extend the vector space ${\bf V}$ over the field ${\mathcal F}$ to include another 
binary operation $\langle \ \mid \ \rangle$, leading to the quartet ${\mathscr H} = \bigl( {\bm V} ,
+ , \cdot , \langle \ \mid \ \rangle \bigr)$. The {\it inner product} $\langle \ \mid \ \rangle$ is
defined as,\footnote{In books on linear algebra, it is common to use the notation 
$\langle \, , \rangle$ for inner product.}
\begin{equation}
    \langle \ \mid \ \rangle :\ {\bm V} \times {\bm V} \ \to \  {\mathcal F} \quad \Longrightarrow \quad  
    \langle {\mathbf u} \mid {\mathbf v} \rangle \ \mapsto \ a,  \quad \quad \forall \  {\mathbf u},  {\mathbf v}  \in {\bm V}, \ a \in {\mathcal F},  
\end{equation} 
and satisfies the following axioms,\footnote{We will ignore $\cdot$ symbol for scalar multiplication.}
\begin{enumerate}
\item $\langle {\mathbf u} \mid {\mathbf u} \rangle \ \ge \ 0 \ {\rm and} \in {\mathbb R}$
\item $\langle {\mathbf u} \mid {\mathbf u} \rangle \ = \ 0 \ {\rm if\ and\ only\ if} \ \mathbf{u} = {\mathbf 0}$
\item $\langle {\mathbf u} + {\mathbf v} \mid {\mathbf w} \rangle = \langle {\mathbf u} \mid {\mathbf w} \rangle + \langle {\mathbf v} \mid {\mathbf w} \rangle $
\item $\langle {\mathbf u} \mid {\mathbf v} + {\mathbf w} \rangle = \langle {\mathbf u} \mid {\mathbf v} \rangle + \langle {\mathbf u} \mid {\mathbf w} \rangle $
\item $\langle a {\mathbf u} \mid {\mathbf v} \rangle =  a^*\, \langle {\mathbf u} \mid {\mathbf v} \rangle $
\item $\langle {\mathbf u} \mid {\mathbf v} \rangle ={\langle {\mathbf v} \mid {\mathbf u} \rangle}^*$
\item $\langle {a \mathbf u} \mid b {\mathbf v} \rangle =  a^* b \langle {\mathbf u} \mid {\mathbf v} \rangle $
\end{enumerate}  
where ${\mathbf w} \in {\bm V}$, $b \in \ {\mathcal F}$, and $*$ in axioms 6 and 7 
denotes the complex conjugate;\footnote{
Mathematicians tend to express axiom 7 as $\langle {a \mathbf u} \mid b {\mathbf v} \rangle =  a b^* \langle {\mathbf u} \mid {\mathbf v} \rangle$. This is 
not an issue as long as we remain consistent with this form used by physicists.} for ${\mathcal F} = {\mathbb R}$ axiom 6 is a symmetry relation. 
Then ${\mathscr H}$ is an {\it inner product (vector) space}.

We define the {\it norm} or {\it length} of a vector ${\mathbf u} \in {\mathscr H}$ as, 
\begin{equation}
|| {\mathbf u} ||= \sqrt{ \langle {\mathbf u} \mid {\mathbf u} \rangle } \label{norm}, 
\end{equation}
which is a real quantity (axiom 1) in ${\mathcal F}$, and has the following properties, 
\begin{itemize}
\item $|| {\mathbf u} || \ge 0$; $|| {\mathbf u} || = 0$ if and only if ${\mathbf u} = 0$
\item $|| a  {\mathbf u} || = | a | \, || {\mathbf u} ||$, where $a \in {\mathcal F}$
and $\mid \dots \mid$ denotes the absolute value  
\item the Cauchy-Schwarz inequality, \[ \bigl| \langle {\mathbf u} \mid {\mathbf v} \rangle \bigr| \le || {\mathbf u} || \ || {\mathbf v} ||, \]
for ${\mathbf v} \in {\mathscr H}$. 
\end{itemize}
Any inner product space is a {\it normed linear space}. 

If, for a sequence of elements $\left\{ {\mathbf u}_1, {\mathbf u}_2, \dots , {\mathbf u}_i, \dots  \right\} \in {\mathscr H}$, the sum,
\begin{equation}
\sum_{i=1}^\infty \ || {\mathbf u}_i ||,
\end{equation}
is absolutely convergent, and all partial sums converge to an element in ${\mathscr H}$, then ${\mathscr H}$ is a {\it complete inner product space} --
it is a {\it Hilbert space}.

Let us define a {\it metric} or {\it distance} function on ${\mathscr H}$ as,
\begin{equation}
d : {\mathscr H} \times {\mathscr H} \to {\mathbb R}_{\ge 0} \quad \Longrightarrow \quad 
\left( {\mathbf u}, {\mathbf v} \right) \mapsto d \left( {\mathbf u}, {\mathbf v} \right) =
|| \left( {\mathbf u}- {\mathbf v} \right) ||, \label{distd}
\end{equation}
with these properties, 
\begin{itemize}
\item $ d \left( {\mathbf u}, {\mathbf v} \right) = 0$ if and only if $ {\mathbf u} = {\mathbf v}$
\item symmetry: $ d \left( {\mathbf u}, {\mathbf v} \right) = d \left( {\mathbf v}, {\mathbf u} \right)$
\item triangle inequality: $ d \left( {\mathbf u}, {\mathbf v} \right) + d \left( {\mathbf v}, {\mathbf w} \right) \le d \left( {\mathbf u}, {\mathbf w} \right)$. 
\end{itemize}
It is intuitively apparent that $d \left( {\mathbf u}, {\mathbf v} \right)$ is a measure of the {\it distance} between two elements in ${\mathscr H}$.
The pair $\left( {\mathscr H}, d \right)$ is referred to as a {\it metric space}.

A sequence of elements $\left\{ {\mathbf u}_1, {\mathbf u}_2, \dots , {\mathbf u}_m, \dots , {\mathbf u}_n, \dots \right\} \in \left( {\mathscr H}, d \right)$ is
a Cauchy sequence if, for all $\epsilon > 0$, there exists a positive integer ${\mathscr N}$ such that for $n, \, m \ge {\mathscr N}$,
$d \left( {\mathbf u}_m, {\mathbf u}_n \right) < \epsilon$. A {\it complete metric space} is one in which every Cauchy sequence converges to
an element in that space. Besides being a complete inner product space, a Hilbert space is also a complete metric space.

\subsubsection{Orthogonality and Orthonormal Basis}

Two elements of ${\mathscr H}$, ${\mathbf u}$ and ${\mathbf v}$, are {\it orthogonal} if $\langle {\mathbf u} \mid {\mathbf v} \rangle = 0$; 
it can also be written as ${\mathbf u} \perp {\mathbf v}$. 
As a consequence, ${\mathbf 0} \in {\mathscr H}$ is orthogonal to all ${\mathbf u} \in {\mathscr H}$. 
If ${\mathbb W}$ is a subspace of ${\mathscr H}$ and ${\mathbb W}^\perp$ is a set of all vectors orthogonal to ${\mathbb W}$, 
\begin{equation}
{\mathbb W}^\perp = \left\{ {\mathbf w} \in {\mathscr H} \mid \langle {\mathbf w} \mid {\mathbf u} \rangle = 0, \ \forall \ {\mathbf u} \in {\mathbb W} \right\}, 
\end{equation}
then ${\mathbb W}^\perp$ is a subspace and ${\mathscr H} = {\mathbb W} \oplus {\mathbb W}^\perp$.
A set of orthogonal vectors where each vector has a unit norm is an {\it orthonormal} set. Every non-empty Hilbert space has an orthonormal basis set. 

\subsubsection{A Note on Metric tensor and Inner Product}
\label{ipdfmt}

In this section, we combine the concepts of metric tensor (Section \ref{sub:metric}) and the inner product (Sections \ref{sub:innpro}).
If ${\mathcal B} = \left\{ {\mathbf e}_1, {\mathbf e}_2, \dots \dots, {\mathbf e}_n \right\}$ is a basis set for vector space ${\bm V}$, then
\begin{equation}
\left< {\mathbf u} \mid {\mathbf v} \right> = \sum\limits_{i, j = 1}^n u^{i*} v^j  \left< {\mathbf e}_i \mid {\mathbf e}_j \right>, 
\end{equation}
where ${\mathbf u}, \ {\mathbf v} \in {\bm V}$ and,
\begin{equation}
{\mathbf u} = \sum\limits_{i = 1}^n u^i {\mathbf e}_i,\quad 
{\mathbf v} = \sum\limits_{i = 1}^n v^i {\mathbf e}_i, \quad
u^i,  v^i  \in {\mathcal F} \ {\rm for} \ i = 1, 2, \dots , n.
\end{equation}
Using the definition of metric tensor in \eqref{met1}, 
\begin{equation}
\left< {\mathbf u} \mid {\mathbf v} \right> = \sum\limits_{i, j = 1}^n u^{i*} v^j  g_{ij}. \label{nmtig}
\end{equation}
If ${\mathcal B}$ is a set of orthonormal vectors, the metric tensor is the identity tensor $g_{ij} = \delta_{ij}$ and, 
\begin{equation}
\left< {\mathbf u} \mid {\mathbf v} \right> = \sum\limits_{i = 1}^n u^{i*} v_i. \label{nmti1}
\end{equation}
The  form in \eqref{nmti1} is similar to the dot product of two vectors in traditional vector analysis.

If ${\mathbf v} = {\mathbf u}$, 
\begin{equation}
\left< {\mathbf u} \mid {\mathbf u} \right> = \sum\limits_{i = 1}^n u^{i*} u_i = \sum\limits_{i = 1}^n |u_i|^2. 
\end{equation}

%% file: bra_ket.tex
\section{Dirac Bra-Ket Notation}
\label{braket}

In the Dirac notation, a vector belonging to a Hilbert space is represented by a {\it ket vector} $\ket{\psi} \in {\mathscr H}$. The quantum
state of a physical system, for example, the spin state of an electron belongs to a Hilbert space and can be denoted by $\ket{\psi}$. The ket vector satisfies the usual
properties related to vectors,
\begin{itemize}
\item if $\ket{\psi}, \ket{\phi} \in {\mathscr H}$ then $\ket{\psi} + \ket{\phi} = \ket{\phi} + \ket{\psi} \in {\mathscr H}$
\item $a \left( b \ket{\psi} \right) =  \left( a b \right) \ket{\psi}$, $\quad \left( a + b \right) \ket{\psi} = a \ket{\psi} + b \ket{\psi}$,\\
$a \left( \ket{\psi} + \ket{\phi} \right) = a \ket{\psi} + a \ket{\phi}$, for $a, b \in {\mathcal F}$.
\item $1 \ket{\psi} = \ket{\psi}$
\item $0 \ket{\psi}$ is the zero ket or {\it null state vector}\footnote{$\ket{0}$ is a state vector, not zero ket or null state vector.}
\end{itemize}

The {\it bra vector}  denoted by $\bra{\phi}$ resides in the dual Hilbert space ${\mathscr H}^*$. The bra is a linear functional
that maps the ket vector onto $\mathcal F$ through the inner product,
\begin{eqnarray}
\bra{\phi} : & {\mathscr H} & \to {\mathcal F} \nonumber \\
& \ket{\psi} & \mapsto \langle \phi | \psi \rangle, \label{dbkt1}
\end{eqnarray}
where $\ket{\psi} \in {\mathscr H}$. There is a one-to-one correspondence between the ket and bra vectors -- for every
$\ket{\psi} \in {\mathscr H}$ there corresponds a unique $\bra{\psi} \in {\mathscr H}^*$, and vice-versa.\footnote{This
follows from the Riesz representation theorem.} The two vectors describe the same state of a system. The relation between them
is,
\begin{equation}
\ket{\psi} = \left( \bra{\psi} \right)^\dagger, \label{dbkt2}
\end{equation}
where $^\dagger$ denotes the Hermitian adjoint or Hermitian conjugate of the bra vector. 
The ket is a column vector while the bra is a row vector.\footnote{In linear algebra terms, the bra is a covector while
the ket is a vector.}
The following are various mappings of interest in the complex Hilbert space.
\begin{enumerate}
\item[(1)]
The mapping,
\[
{\mathscr H} ^* \times {\mathscr H} \to {\mathbb C}  \Longrightarrow \Bigl( \bra{\phi}, \, \ket{\psi} \Bigr) \mapsto \braket{\phi | \psi} \quad  
 {\rm where} \ \bra{\phi} \in {\mathscr H}^* \ {\rm and} \ \ket{\psi} \in {\mathscr H},  
\]
leads to the inner product.

\item[(2)]
In contrast,
\[
{\mathscr H}  \times {\mathscr H}^* \to {\mathscr K}  \Longrightarrow \Bigl( \ket{\psi}, \, \bra{\phi} \Bigr) \mapsto \ket{\psi} \bra{\phi},
\]
is an {\it outer product}, or a {\it linear operator}. Here, ${\mathscr K}$ is another, different, Hilbert space with ${\rm dim} ( {\mathscr K} ) = \bigl( {\rm dim}
( {\mathscr H}) \bigr)^2$. For $\ket{\zeta} \in {\mathscr K}$,
\[
\Bigl( \ket{\psi} \bra{\phi} \Bigr) \, \ket{\zeta} =  \ket{\psi}  \Bigl( \bra{\phi} \,  \ket{\zeta} \Bigr) = 
\braket{{\phi} | {\zeta}} \ket{\psi},
\]
illustrating that $\ket{\psi} \bra{\phi}$ is an operator.

\item[(3)]
The mapping of the tensor product of two different Hilbert spaces ${\mathscr H}_1$ and ${\mathscr H}_2$, is defined as,
\[
 f: {\mathscr H}_1 \otimes {\mathscr H}_2 \to {\mathscr K} \ \Rightarrow \ \Bigl( \ket{\phi}, \ket{\psi} \Bigr) \mapsto \ket{\phi} \otimes \ket{\psi}
\]
where ${\mathscr K}$ is another Hilbert space with ${\rm dim} ( {\mathscr K} ) = \bigl( {\rm dim}
( {\mathscr H}_1) \bigr) \times \bigl( {\rm dim} ( {\mathscr H}_2) \bigr)$, $\ket{\phi} \in {\mathscr H}_1$, 
$\ket{\psi} \in {\mathscr H}_2$, and $\ket{\phi} \otimes \ket{\psi} \in {\mathscr K}$. It is usual to drop the symbol for the tensor product
and replace $\ket{\phi} \otimes \ket{\psi}$ by $\ket{\phi} \ket{\psi}$.\newline
The outer product of vectors in two different Hilbert spaces is defined in the same way as in (2) above.
However, we cannot define an inner product with respect to two different Hilbert spaces. 
\end{enumerate}

\subsection{\textbf {\textit{Linear Operators}}}
\label{sub:linearops}

In the previous section we stated that $\ket{\psi} \bra{\phi}$ is a linear operator. Here we will formalize the definition.
A map,
\begin{equation}
{\mathcal T}: {\mathscr H} \to {\mathscr H} \ \Rightarrow \ {\ket{\psi}} \mapsto {\mathcal T} \left( \ket{\psi} \right) \in {\mathscr H},
\end{equation}  is a { \it linear operator} if and only if,\footnote{Recall that the difference between a linear map and a linear operator is 
that the former is a map between any two vector spaces while the latter is a map between the same vector space.}
\begin{equation}
\mathcal{T} \left( a \ket{\psi} + b \ket{\phi} \right) = a {\mathcal T} \ket{\psi} + b {\mathcal T} \ket{\phi}, \quad 
\ket{\psi}, \ket{\phi} \in {\mathscr H}, \ {\rm and} \ \ a, b \in {\mathbb C}.
\end{equation}
It is straightforward to show that the outer product mentioned in the previous section is a linear operator.
If ${\rm dim} \left( {\mathscr H} \right) = n$, then an operator in this Hilbert space is a complex rank 2 tensor, or  a matrix, with
$n \times n$ elements. 

If, in addition, there is a zero operator ${\bm 0} \in {\mathscr H}$ and an identity operator ${\mathcal I} \in {\mathscr H}$
such that,
\begin{equation}
{\bm 0} \ket{\psi} = 0, \quad {\mathcal I} \ket{\psi} = \ket{\psi}, \quad \forall \ \ \ket{\psi} \in {\mathscr H},
\end{equation}
then the space of all linear operators is a vector space.
Consider a set of orthonormal basis vectors $\ket{e_i}$ that span ${\mathscr H}$ with  $\braket{e_j \mid  e_i} = \delta_{ij}$ for
$i, j = 1, 2, \dots n$.  Any ket vector can be expressed as a linear sum over the basis set,
\begin{equation}
\ket{\psi} = \sum_{i = 1}^n \ c_i \ket{e_i} \label{lo1}
\end{equation}
where $c_i \in {\mathbb C}$. The inner product of $\ket{\psi}$ with $\bra{e_j}$ is,
\begin{equation}
\braket{e_j \mid \psi} = \sum_{i = 1}^n \ c_i \braket{e_j \mid e_i} = c_j. \label{lo2}
\end{equation}
Combining \eqref{lo1} and \eqref{lo2},
\begin{equation}
\ket{\psi} = \sum_{i = 1}^n \ \braket{e_i \mid \psi} \ket{e_i} = \sum_{i = 1}^n \ \ket{e_i} \braket{e_i \mid \psi} =
\sum_{i = 1}^n \ \bigl( \ket{e_i} \bra{e_i} \bigr)\,  \ket{\psi} \label{lo3}.
\end{equation}
Since \eqref{lo3} is valid for any $\ket{\psi} \in {\mathscr H}$, we obtain the identity operator,
\begin{equation}
{\mathcal I} = \sum_{i = 1}^n \  \ket{e_i} \bra{e_i}. \label{lo4}
\end{equation}
By definition, ${\mathcal T} \ket{e_i} \in {\mathscr H}$ is a ket vector. Thus,
\begin{equation}
{\mathcal T} \ket{e_i} = \sum_{k = 1}^n \ T^k_i \ket{e_k}, \label{lo5}
\end{equation}
where $T^k_i \in {\mathbb C}$. It follows that,
\begin{equation}
\bra{e_j} {\mathcal T} \ket{e_i} = \sum_{k = 1}^n \ T^k_i \braket{e_j \mid e_k} = \sum_{k = 1}^n \ T^k_i \delta^j_k
= T^j_i,  \label{lo6}
\end{equation}
giving the elements of the operator ${\mathcal T}$ in the chosen basis set. 
From \eqref{lo4} and \eqref{lo6},
\begin{equation}
{\mathcal T} = \sum_{i, j = 1}^n \ T^i_j \ket{e_i} \bra{e_j}. \label{lo7}
\end{equation}

\paragraph{\bf Hermitian Conjugate of an Operator}
 
Since ${\mathcal T} \ket{\psi} \in {\mathscr H}$ for any $\ket{\psi} \in {\mathscr H}$, one of the properties of inner product leads to,
\begin{equation}
\bra{\phi} {\mathcal T} \ket{\psi} = \bra{\psi} {\mathcal T}^{\dagger} \ket{\phi}^*,  \quad \ket{\phi} \in {\mathscr H}.
\end{equation}
Consequently, following \eqref{lo6},
\begin{equation}
\bra{e_i} {\mathcal T}^{\dagger} \ket{e_j}^* =  {{T^i}_j^*}.
\end{equation}
The matrix representation of ${\mathcal T}^\dagger$ is obtained by taking the complex conjugate of the transpose of ${\mathcal T}$.
An operator is {\it Hermitian} or {\it self-adjoint} if ${\mathcal T}^\dagger = {\mathcal T}$. As a result, in a matrix representation of ${\mathcal T}$,
${{T^i}_j^*} = T^j_i$ -- the diagonal terms are real.

\paragraph{\bf Unitary Operator}

For a linear operator ${\mathcal U}: {\mathscr H} \to {\mathscr H}$, let $\ket{\phi} = {\mathcal U} \ket{\psi}$ where
$\ket{\phi}, \ket{\psi} \in {\mathscr H}$. Then, $\bra{\phi} = \bra{\psi} {\mathcal U}^\dagger$
and,
\begin{equation}
\braket{\phi \mid \phi} = \bra{\psi} {\mathcal U}^\dagger {\mathcal U} \ket{\psi}. \label{u1}
\end{equation}
${\mathcal U}$ is a {\it unitary operator} if and only if,
\begin{equation}
{\mathcal U}^\dagger {\mathcal U} = {\mathcal I} \Longrightarrow {\mathcal U}^\dagger = {\mathcal U}^{-1}. \label{u2}
\end{equation}
Then, from \eqref{u1}, $|| \phi ||^2 = || \psi ||^2$. Thus, a unitary operator ${\mathcal U}$ transforms any vector in
${\mathscr H}$ in such a way as to preserve the norm.

\paragraph{\bf Projection Operator}

For the orthonormal basis vectors $\ket{e_i}$, $i = 1, 2, \dots , n$, mentioned in Section \ref{sub:linearops}, define
an operator,
\begin{equation}
{\mathcal P}_i = \ket{e_i} \bra{e_i},
\end{equation}
which has the properties,
\begin{equation}
\mathcal P_i{^2} = \ket{e_i} \underbrace{\braket{e_i \mid e_i}}_{=\ 1} \bra{e_i} = \ket{e_i} \bra{e_i} = \mathcal P_i,
\quad \quad \quad \sum_{i=1}^n \ \mathcal P_i = {\mathcal I}.
\end{equation}
For $\ket{\psi}$ defined in \eqref{lo1},
\begin{equation}
\mathcal P_i \ket{\psi} = \mathcal P_i \sum_{j=1}^n \ c_j \ket{e_j} = \sum_{j=1}^n \ c_j \ket{e_i} 
\underbrace{\braket{e_i \mid e_j}}_{\delta_{ij}} = c_i \ket{e_i}.
\end{equation}
Thus, the operator $\mathcal P_i$ determines the component of $\ket{\psi}$ in the $\ket{e_i}$ direction; 
$\mathcal P_i$ is referred to as a {\it projection operator}. 

We can generalize the concept of a projection operator by defining,
\begin{equation}
\mathcal P = \ket{\psi} \bra{\psi},
\end{equation}
with the normalization $\braket{\psi \mid \psi} = 1$. Clearly, $\mathcal P{^2} = \mathcal P $. If $\ket{\phi} \in {\mathscr H} $
is another vector, then,
\begin{equation}
\mathcal P \ket{\phi} = \braket{\psi \mid \phi} \ket{\psi},
\end{equation}
is a projection of $\ket{\phi}$ along $\ket{\psi}$.

%% file: quantum_postulates.tex
\section{Postulates of Quantum Mechanics}
\label{pqm} 

There are four basic postulates of quantum physics which guide the development of quantum computers.
We will state the essential aspects of these postulates without getting into finer details.

\paragraph{\bf Postulate 1}

The quantum state of a particle is described by a state vector $\ket{\psi}$ which is an element belonging to a complex
Hilbert space ${\mathscr H}$ -- the state space of quantum theory. 

\paragraph{\bf Postulate 2}

Associated with each dynamical variable ${\mathrm A}$, such as energy, position, momentum of a particle, there is a unique Hermitian
(self-adjoint) operator $\widehat{\mathit A}$ which operates on the states in ${\mathscr H}$.  
The normalized eigenvectors of $\widehat{\mathit A}$,  $\ket{e_i}$, form a complete set which spans
${\mathscr H}$.\footnote{We assume that  ${\rm dim} \left( {\mathscr H} \right) = n$, and $i = 1, 2, \dots , n$.} Each of the corresponding
eigenvalues $\lambda_i$ is a possible value of the observable ${\mathrm A}$.\footnote{
$\widehat{\mathit A} \ket{e_i} = \lambda_i \ket{e_i}$.}

\paragraph{\bf Postulate 3}

Any state can be expanded in terms of the eigenvectors of a Hermitian operator,
\begin{equation}
\ket{\psi} = \sum_{j=1}^n \ c_j \ket{e_j},
\end{equation}
where $c_j \in {\mathbb C}$, $j = 1, 2, \dots , n$.
The probability ${\mathrm P}$ 
of observing a particular eigenvalue $\lambda_i$ is obtained using the projection operator $\mathcal P_i = \ket{e_i} \bra{e_i}$,
\begin{equation}
{\mathrm P} \left( \lambda_i \right) = \braket{ \psi \mid \mathcal P_i \mid \psi } = 
\sum_{j, k = 1}^n \ c_k^* c_j \underbrace{\braket{e_k \mid e_i} }_{\delta_{ik}} \underbrace{\braket{e_i \mid e_j} }_{\delta_{ij}} = 
\left| c_i \right|^2.
\end{equation}
When a measurement of $\ket{\psi}$ yields $\lambda_i$ for ${\mathrm A}$, the state $\ket{\psi}$ collapses to the eigenstate $\ket{e_i}$.

\paragraph{\bf Postulate 4}

The time evolution of the state vector $\ket{\psi (t)}$ is given by the Schr\"odinger equation,
\begin{equation}
i \hbar \frac{d}{dt} \ket{\psi(t)} = {\mathbf H} \left( t \right) \, \ket{\psi(t)}, \label{sch1}
\end{equation}
where $\hbar = h / 2 \pi$, $\ket{\psi (t) }$ is the state vector at time $t$, and
${\mathbf H}$ is a Hermitian operator, also referred to as a {\it Hamiltonian}. It should be noted that
${\mathbf H}$ has the same units as energy. The evolution equation is valid for a closed quantum system -- in other words,
the quantum system of interest is not a subset of a larger quantum system. 
From \eqref{sch1},
\begin{align}
\frac{d}{dt} \braket{\psi (t) \mid \psi (t)} & = \left( \frac{d}{dt} \bra{\psi (t)} \right) \ket{\psi (t)}
+ \bra{\psi (t)} \left( \frac{d}{dt} \ket{\psi (t)} \right) \nonumber \\
& =
\left\{ \bra{\psi (t)} \left( \frac{1}{i \hbar} {\mathbf H} \right)^\dagger \right\} \ket{\psi (t)} +
\bra{\psi (t)} \left\{ \left( \frac{1}{i \hbar} {\mathbf H} \right)  \ket{\psi (t)} \right\} \nonumber \\
& = \frac{1}{i \hbar} \bra{\psi (t)} \left( - {\mathbf H}^\dagger + {\mathbf H} \right) \ket{\psi (t)} = 0, \label{sch2}
\end{align}
since ${\mathbf H}$ is a Hermitian operator. The Schr\"odinger equation preserves the norm during
the time evolution of the state vector. As shown above, unitary operators also preserve the norm.
Suppose there exists a unitary operator ${\mathcal U}$ which maps $\ket{\psi (t_0)}$ at time $t_0$
to $\ket{\psi (t)}$ at time t,
\begin{equation}
{\mathcal U}\left( t, t_0 \right): \ket{\psi (t_0)} \to \ket{\psi (t)} \quad \Rightarrow \quad \ket{\psi(t_0)} \mapsto \ket{\psi(t)} = {\mathcal U} \left( t, t_0 \right) 
\ket{\psi(t_0)}. \label{sch3}
\end{equation}
Then,
\begin{equation}
i \hbar \frac{d}{dt} \ket{\psi (t)} =i \hbar \left\{ \frac{d}{dt} {\mathcal U} \left( t, t_0 \right) \right\} \ \ket{\psi (t_0)}. \label{sch4}
\end{equation}
From \eqref{sch4}, and \eqref{sch1},
\begin{equation}
i \hbar \left\{ \frac{d}{dt} {\mathcal U} \left( t, t_0 \right) \right\} \ \ket{\psi (t_0)} = 
 {\mathbf H} \left( t \right) \, {\mathcal U} \left( t, t_0 \right)  \ \ket{\psi (t_0)}. \label{sch5}
\end{equation}
Since this equation is valid for any $\ket{ \psi (t_0)}$, we conclude that,
\begin{equation}
i \hbar  \frac{d}{dt} {\mathcal U} \left( t, t_0 \right) = 
 {\mathbf H} \left( t \right) \, {\mathcal U} \left( t, t_0 \right). \label{sch6}
\end{equation}
When the Hamiltonian does not depend on time, \eqref{sch6} can be integrated and we obtain,
\begin{equation}
{\mathcal U} \left( t, t_0 \right)  = \exp \left\{ \frac{i}{\hbar} {\mathbf H} \left( t - t_0 \right) \right\}. \label{sch7}
\end{equation}
It follows that Hermitian operators are generators of unitary operators.

%% file: bits_qubits.tex
\section{Basic Aspects of Quantum Computing}
\label{sec:basicqc}

In this section, we will discuss some of the differences between classical computing and quantum computing.
The discussion will also illustrate aspects of quantum computing which fuel the passion for developing quantum
computers.

\subsection{\textbf{\textit{Bits Versus Qubits: Linear Superposition}}}

In classical computers, the smallest unit of data storage is a bit. A bit takes on a single binary
value -- either 0 (off) or 1 (on). A two bit system can store data that, for example, is equivalent
to either 0, 1, 2, or 3 in the decimal system. Three bits can represent integers ranging from 0 to
7 in the decimal system. Generalizing, $n$ bits can store {\it any one} integer in the range $\left[ 0,  2^n-1 \right]$.
While bits ``live'' in a discrete binary space, the building blocks of a quantum computer {\it qubits} 
(quantum bits) belong to a two-dimensional Hilbert space ${\mathscr H}_2$ over a field of complex numbers.
Any two orthonormal vectors that span ${\mathscr H}_2$ can be loosely thought of as being equivalent
to the two states of a bit. However, there are important differences that are worth noting.

Let us choose the two orthonormal vectors that span ${\mathscr H}_2$ to be,
\begin{equation}
\ket{0} = \begin{pmatrix}  1 \\ 0 \end{pmatrix}, \quad \ket{1} = \begin{pmatrix}  0 \\ 1 \end{pmatrix}. \label{ba1}
\end{equation}
These two vectors can be considered as representing the two spin directions of an electron, or the two polarizations of
an electromagnetic wave propagating in vacuum. Note that $\braket{0 \mid 1} = \braket{ 1 \mid 0} = 0$ and
$\braket{0 \mid 0} = \braket{ 1 \mid 1} = 1$. The two vectors in \eqref{ba1} are sometimes referred to as
the {\it computational basis} states. Any state vector $\ket{\psi} \in {\mathscr H}_2$ has the form,
\begin{equation}
\ket{\psi} = a_1 \ket{0} + b_1 \ket{1}, \label{ba2}
\end{equation}
where $a_1, b_1 \in {\mathbb C}$. In other words, a qubit is a {\it linear superposition} of the two basis states that
span ${\mathscr H}_2$.
In contrast, a classical bit can only be in one of two discrete states. The ability of a qubit to retain a linear superposition
of two states distinguishes quantum computing from classical computing.

There is a difference between the superposition principle in linear classical physics and in linear quantum physics. For example,
the electrostatic force on a test charge by an accumulation of other field charges surrounding it, is a linear sum of the electrostatic force
due to each field charge on the test charge. Since the electrostatic force between any pair of charges is a well-defined force, so is the
linear superposition. In comparison, $\ket{0}$ and $\ket{1}$ correspond to physical entities, for example, spin up or spin down of an
electron, but the superposition \eqref{ba2} is not a physically observable state. It prescribes a probability to observing any one of the 
two states. Consequently, following Postulate 3, the probability of observing the state $\ket{0}$ is,
\begin{equation}
P \left( \ket{0} \right) = \left| \braket{0 \mid \psi} \right|^2 = \left| a_1 \braket{0 \mid 0} + b_1 \braket{0 \mid 1} \right|^2 = \left| a_1 \right|^2,
\label{ba3}
\end{equation}
while that of observing the state $\ket{1}$ is $\left| b_1 \right|^2$. Since the system can be in only one of these two states,
\begin{equation}
\braket{\psi \mid \psi} = 1 \quad \Rightarrow \quad \left| a_1 \right|^2 + \left| b_1 \right|^2 = 1, \label{ba4}
\end{equation}
as expected. Thus, when subject to a measurement, a qubit contains one classical bit of information. It will be either in state $0$ (off) with
probability $\left| a_1 \right|^2$ or in state $1$ (on) with probability $ 1 - \left| a_1 \right|^2$.

The state space for a two qubit system is the tensor product ${\mathscr H}_2^{(1)} \otimes {\mathscr H}_2^{(2)}$, where the superscript
distinguishes the two isomorphic Hilbert spaces. Without loss of generality, we assume that the basis vectors for the two Hilbert
spaces are the same. Consider the two state vectors,
\begin{equation}
\ket{\psi}_1 = a_1 \ket{0} + b_1 \ket{1} \in {\mathscr H}_2^{(1)}, \quad 
\ket{\psi}_2 = a_2 \ket{0} + b_2 \ket{1} \in {\mathscr H}_2^{(2)}, \label{ba5}
\end{equation}
with $ a_1, b_1, a_2, b_2 \in {\mathbb C}$. Each qubit state is normalized,
\begin{equation}
\left| a_1 \right|^2  + \left| b_1 \right|^2  = 1, \quad \left| a_2 \right|^2  + \left| b_2 \right|^2  = 1. \label{ba6}
\end{equation}
Let us
denote the two-qubit state $\ket{\psi}_1 \otimes \ket{\psi}_2 \in {\mathscr H}_2^{(1)} \otimes {\mathscr H}_2^{(2)}$ by $\ket{\psi_1 \psi_2}$. Then,
\begin{equation}
\ket{\psi_1 \psi_2} = a_1 a_2 \ket{0 0} + a_1 b_2 \ket{0 1} + b_1 a_2 \ket{1 0} + b_1 b_2 \ket{1 1}, \label{ba7}
\end{equation}
where the basis vectors are,\footnote{$\ket{0} \otimes \ket{0} = \ket{ 0 0}$, etc.}
\begin{equation}
\ket{0 0} = \begin{pmatrix} 1 \\ 0 \\ 0 \\ 0 \end{pmatrix},  \quad
\ket{0 1} = \begin{pmatrix} 0 \\ 1 \\ 0 \\ 0 \end{pmatrix},  \quad
\ket{1 0} = \begin{pmatrix} 0 \\ 0 \\ 1 \\ 0 \end{pmatrix},  \quad
\ket{1 1} = \begin{pmatrix} 0 \\ 0 \\ 0 \\ 1 \end{pmatrix}. \label{ba8}
\end{equation}
The normalization $\braket{ \psi_1 \psi_2\mid \psi_1 \psi_2} = 1$ imposes the constraint,
\begin{equation}
\left| a_1 a_2 \right|^2 + \left| a_1 b_2 \right|^2 + 
\left| b_1 a_2 \right|^2 + \left| b_1 b_2 \right|^2  = 1. \label{ba9}
\end{equation} 
However, the three normalizations in \eqref{ba6} and \eqref{ba9} are not independent; \eqref{ba9} can be obtained
from the two equations in \eqref{ba6}.

From \eqref{ba7}, we note that a two qubit system is a superposition of four basis states.
A classical two bit system can be in only one of four possible states. Generalizing, an $n$-bit classical system represents only one of $2^n$ distinct states,
while a $n$-qubit system is a superposition of $2^n$ states. Consequently, it is expected that quantum computers will be able to manipulate
exponentially larger data sets compared to classical computers.

\subsection{\textbf{\textit{Entanglement of Qubits}}}

The two-qubit state \eqref{ba7} results from a tensor product of two separate Hilbert spaces -- each qubit ``residing'' in its own two-dimensional
Hilbert space. Intrinsically, we treat each qubit as an isolated physical system. However, the qubits can be set up to interact with each other such
that, in the language of linear algebra, the combined system is a state in a four dimensional Hilbert space ${\mathscr H}_4$. We can still use the
same formalism as in the preceding subsection. The basis vectors of ${\mathscr H}_4$ are given in \eqref{ba8}, and any state of the combined
system has the form,
\begin{equation}
\ket{\psi} = \alpha_1 \ket{0 0} + \alpha_2 \ket{0 1} + \alpha_3 \ket{1 0} + \alpha_4 \ket{1 1}, \label{ba10}
\end{equation}
where $\left\{ \alpha_1, \alpha_2, \alpha_3, \alpha_4 \right\} \in {\mathcal F}$, and
\begin{equation}
\left| \alpha_1 \right|^2 + \left| \alpha_2 \right|^2 + 
\left| \alpha_3 \right|^2 + \left| \alpha_4 \right|^2  = 1. \label{ba11}
\end{equation}
Following the same steps as in Section \ref{subsub:tensorprod},
we illustrate the difference between a state vector of two isolated qubits \eqref{ba7}  and the state vector of a combined two-qubit system \eqref{ba10}.
If we set, for example, $\alpha_1 = 0$, there does not exist any set $\left\{ a_1, b_1, a_2, b_2 \right\} \in {\mathcal F}$ for which
\eqref{ba7} leads to the same result as \eqref{ba10}. This is not surprising since ${\mathscr H}_2 \otimes {\mathscr H}_2 \subset {\mathscr H}_4$.
Consequently, there exist state vectors in ${\mathscr H}_4$ which are not part of the vector space ${\mathscr H}_2 \otimes {\mathscr H}_2$. 
Stated differently, there are state vectors in ${\mathscr H}_4$ which cannot be factored as a tensor product of states in ${\mathscr H}_2$.
Such states are {\it irreducible} and are referred to as {\it entangled} states.
For a two qubit system, there are four {\it maximally entangled} states -- the {\it Bell} states -- which are elements in ${\mathscr H}_4$,\footnote{
We will leave it as an exercise for the reader to ascertain the meaning of {\it maximally entangled} Bell states.}
\begin{equation}
\ket{\Psi}_{B \pm} = \frac{1}{\sqrt{2}} \bigl( \ket{0 0} \pm \ket{1 1} \bigr), \quad
\ket{\Phi}_{B \pm} = \frac{1}{\sqrt{2}} \bigl( \ket{0 1} \pm \ket{1 0} \bigr).
\label{ba12}
\end{equation}
The four Bell states $\ket{\Psi}_{B \pm}$ and $\ket{\Phi}_{B \pm}$ can also be chosen as the orthonormal basis set spanning ${\mathscr H}_4$.
It is worth noting that the Bell states are entangled, but not all entangled states are Bell states. 

\paragraph{\bf Logic Gates for Entanglement}

There are two quantum logic gates -- the controlled NOT (${\rm CNOT}$)  gate and the Hadamard (${\rm H}$) gate -- which connect the
entangled Bell states with the basis states in \eqref{ba8}. For our purposes, it is sufficient to give the matrix forms that represent the two logic gates,
\begin{equation}
{\rm CNOT} = \begin{bmatrix} 
1 & 0 & 0 & 0 \\
0 & 1 & 0 & 0 \\
0 & 0 & 0 & 1 \\
0 & 0 & 1 & 0 
\end{bmatrix}, \quad \quad
{\rm H} = \frac{1}{\sqrt{2}} \, 
\begin{bmatrix} 
1 & 1 \\
1 & -1 
\end{bmatrix}. \label{ba13}
\end{equation}
It is useful to note that ${\rm CNOT}$ and ${\rm H}$ are unitary matrices and their own inverses. Also,
${\rm CNOT}$ operates on a two-qubit state, while ${\rm H}$ operates on a single qubit. 
Thus,
\begin{equation}
{\rm CNOT} \ket{0 0} = \ket{0 0},\ 
{\rm CNOT} \ket{0 1} = \ket{0 1},\ 
{\rm CNOT} \ket{1 0} = \ket{1 1},\ 
{\rm CNOT} \ket{1 1} = \ket{1 0}, \nonumber \label{ba14}
\end{equation}
\begin{equation}
{\rm H} \ket{0} = \frac{1}{\sqrt{2}} \bigl( \ket{0} + \ket{1} \bigr), \quad {\rm H} \ket{1} = \frac{1}{\sqrt{2}} \bigl( \ket{0} - \ket{1} \bigr).
\label{ba15}
\end{equation}
If we first operate on \eqref{ba12} with ${\rm CNOT}$ and then operate on the first qubit with $H$,\footnote{We will get the same results
if we operate on the first qubit with ${\rm H}$ and then with ${\rm CNOT}$ on each two-qubit state.}
\begin{eqnarray}
\ket{\Psi}_{B\pm} & \xleftrightarrow{{\rm CNOT}} &\frac{1}{\sqrt{2}} \bigl( \ket{0 0} \pm \ket{1 0} \bigr) \nonumber \\
& \xleftrightarrow{\ \ {\rm H} \ \ } & \frac{1}{2}
\left[ \ket{ 0 0} + \ket{ 1 0} \pm \bigl( \ket{ 0 0} - \ket{ 1 0} \bigr) \right] \ = \ \begin{cases} +: & \ket{0 0}\\ -: & \ket{1 0} \end{cases}, \\
\ket{\Phi}_{B\pm} & \xleftrightarrow{{\rm CNOT}} &\frac{1}{\sqrt{2}} \bigl( \ket{0 1} \pm \ket{1 1} \bigr) \nonumber \\
& \xleftrightarrow{\ \ {\rm H} \ \ } & \frac{1}{2}
\left[ \ket{ 0 1} + \ket{ 1 1} \pm \bigl( \ket{ 0 1} - \ket{ 1 1} \bigr) \right] \ = \ \begin{cases} +: & \ket{0 1}\\ -: & \ket{1 1} \end{cases}. \label{ba16}
\end{eqnarray}
As the arrows indicate, the process is completely reversible. Thus, we can connect irreducible Bell states to the
basis set for the ${\mathscr H}_2 \otimes {\mathscr H}_2$ Hilbert space. For example, if a measurement gives the state $\ket{0 1}$,
then, with complete certainty, we know  that the entangled state of two qubits is ${\Phi}_{B+}$.

\paragraph{\bf Computational Advantage of Entanglement}

A single qubit $\ket{\psi} = a \ket{0} + b \ket{1}$, $ a, b \in {\mathbb C}$, is defined by two complex numbers or, equivalently,
an ordered set of two pairs of real numbers (Section \ref{sub:fld}). Thus, the set of all coefficients $a$ and $b$ is isomorphic to ${\mathbb R}^4$.
The normalization constraint $\left| a \right|^2  + \left| b \right|^2 = 1$ reduces the dimension to ${\mathbb R}^3$.\footnote{The dimension
can be reduced to ${\mathbb R}^2$ by ignoring a global phase which does not affect the physical state. Nevertheless,
it does not impact our overall line of reasoning.} In other words, the parameter space of a single qubit is three-dimensional real space.
The set of all coefficients in the reducible form \eqref{ba7} of two qubits is isomorphic to ${\mathbb R}^6$ since there are two normalization
constraints. In general, for a reducible state formed by the tensor product of $n$ qubits, the set of all coefficients is isomorphic to
${\mathbb R}^{3n}$.

The parameter space of entangled states formed by two qubits \eqref{ba10}
is isomorphic to ${\mathbb R}^7$ as there is only
one normalization constraint \eqref{ba11}. For $n$ qubits in an entangled state,\footnote{$ \left\{ n \in {\mathbb N} \mid n \ge 2 \right\}$.} subject to
a single normalization constraint, the parameter space is isomorphic to ${\mathbb R}^{2^{n+1} - 1}$. Thus, entangling of $n$ qubits
increases the state space from $3n$ real dimensions to $2^{n+1}-1$ dimensions; that is, the state space of entangled, or irreducible, states
grows exponentially while that of the reducible states grows linearly with the number of qubits.

Another advantage of entangled states can be illustrated by a simple example. Consider a two qubit reducible state,
\begin{equation}
\ket{\psi} = \left\{ \frac{1}{\sqrt 2} \bigl( \ket{0} + \ket{1} \bigr) \right\} \otimes \left\{ \frac{1}{\sqrt 2} \bigl( \ket{0} + \ket{1} \bigr) \right\}
= \frac{1}{2} \bigl( \ket{0 0} + \ket{0 1} + \ket{1 0} + \ket{1 1} \bigr). \label{ba17}
\end{equation}
Suppose we measure the first qubit to be in state $\ket{0}$. The second qubit remains in its superposition of states and the probability
of measuring the second qubit to be $\ket{0}$ is $1/2$. In comparison, for an entangled Bell state $\ket{\Psi}_{B\pm}$ from 
\eqref{ba12}, if we measure the first qubit to be in state $\ket{0}$, then the second qubit is also in state $\ket{0}$ with probability 1.
Thus, not only is the state space of entangled qubits exponentially larger than for reducible states, the access to information is 
also more efficient. Entanglement plays an essential role in quantum computations and in the speedup of quantum computers.

%% file: maxwell.tex
\section{Maxwell Equations and Their Covariant Form}
\label{sec:maxwellco}

The macroscopic Maxwell equations in a medium are usually expressed as,
\begin{eqnarray}
\nabla \cdot {\mathbf D} \left( {\mathbf r}, t \right) & = & \rho_e \left( {\mathbf r}, t \right), \label{1.1}\\
\nabla \cdot {\mathbf B} \left( {\mathbf r}, t \right) & = & 0, \label{1.2} \\
\nabla \times {\mathbf E} \left( {\mathbf r}, t \right) & = & - \frac{\partial}{\partial  t} {\mathbf B} \left( {\mathbf r}, t \right), \label{1.3}\\
\nabla \times {\mathbf H} \left( {\mathbf r}, t \right) & = &  \frac{\partial}{\partial  t} {\mathbf D} \left( {\mathbf r}, t \right) +
 {\mathbf j}_e \left( {\mathbf r}, t \right),  
\label{1.4}
\end{eqnarray} 
where $\rho_e \left( {\mathbf r}, t \right)$ and ${\mathbf j}_e\left( {\mathbf r}, t \right)$ are externally applied electric 
charge and current densities, respectively. 
Additionally, ${\mathbf E}$ is the electric field, ${\mathbf D}$ is the displacement electric field, ${\mathbf B}$ is the magnetic induction or magentic field,
and ${\mathbf H}$ is the magnetic intensity. We will refer to \eqref{1.1} as Gauss' law, \eqref{1.2} as Gauss' law for magnetism, \eqref{1.3} as 
the Faraday equation, and \eqref{1.4} as the Ampere-Maxwell equation.

The system of equations \eqref{1.1}-\eqref{1.4} is not determinate; we have four vector equations with six vector unknowns. 
The two additional conditions needed to make the system determinate are the {\it constitutive relations},
\begin{equation}
{\mathbf D} = \epsilon_0 {\mathbf E} + {\mathbf P}, \quad {\mathbf B} = \mu_0 \left( {\mathbf H} + {\mathbf M} \right),
\end{equation}
where ${\mathbf P}$ and ${\mathbf M}$ are the polarization and magnetization densities, respectively, of the medium,
and $\epsilon_0$ and $\mu_0$ are the permittivity and permeability, respectively, of vacuum. The response of the medium
to ${\mathbf E}$ and ${\mathbf B}$ determines ${\mathbf P}$ and ${\mathbf M}$, and requires physics modeling that extends beyond
Maxwell equations. In vacuum ${\mathbf P} = 0$ and ${\mathbf M} = 0$, while for a dielectric medium, such as a plasma, ${\mathbf M} = 0$.

If we take the divergence of \eqref{1.4}  and the time derivative of \eqref{1.1}, we obtain the charge conservation equation,
\begin{equation}
\frac{\partial}{\partial t} \rho_e \left( {\mathbf r}, t \right) + \nabla \cdot {\mathbf j}_e \left( {\mathbf r}, t \right) = 0. \label{1.7}
\end{equation}
If we impose charge conservation, the divergence of \eqref{1.4} yields,
\begin{equation}
\frac{\partial}{\partial t} \left[  \nabla \cdot {\mathbf D} \left( {\mathbf r}, t \right)\right] = 0. \label{1.5}
\end{equation}
The divergence of \eqref{1.3} leads to,
\begin{equation}
\frac{\partial}{\partial t} \left[  \nabla \cdot {\mathbf B} \left( {\mathbf r}, t \right)\right] = 0. \label{1.6}
\end{equation}
Thus, if \eqref{1.1} and \eqref{1.2} are satisfied at time $t=0$, \eqref{1.5} and \eqref{1.6} ensure that \eqref{1.1} and \eqref{1.2} are
satisfied for all times. In effect, the two Gauss' laws \eqref{1.1} and \eqref{1.2} are
initial conditions for the Faraday and Ampere-Maxwell equations. 
However, this statement is only true in the continuum limit. In the
discrete limit, appropriate for computations, this statement is not necessarily valid. Consequently, it is necessary to solve the four
Maxwell equations simultaneously.

\subsection{\textbf{\textit{Four Vectors, Minkowski Space, and Special Relativity}}}

The electromagnetic fields are functions of space ${\mathbf r}$ and time $t$ and, thereby, residing in a four-dimensional space-time vector space. 
In this space a vector is referred to as a 4-vector with its first component being time-like and the other three components being space-like.
The space-time 4-vector in terms of its components ${\mathbf x} = \left( x^0, x^1, x^2, x^3 \right)$ 
is a contravariant vector, with $ x^0 = ct $, and $ x^1, x^2, x^3 $ being the
usual Cartesian coordinates $x, y, z$, respectively. The corresponding  vector space is the {\it Minkowski space}. We define a bilinear
functional in real Minkowski vector space ${\mathscr M}$,
\begin{equation}
\eta : {\mathscr M} \times {\mathscr M} \to {\mathbb R}. \label{min1}
\end{equation}
From the discussion in Sections \ref{subs:multilinear} and \ref{sub:metric},
\begin{equation}
\eta : {\mathscr M}^* \otimes {\mathscr M}^* \to {\mathbb R}, \label{min2}
\end{equation}
is a covariant tensor -- a metric tensor -- of rank 2. It is usually referred to as {\it Minkowski metric} 

In what follows, we will adopt the convention that Greek 
indices (superscripts or subscripts) are $0$, $1$, $2$, or $3$,  while Latin indices are $1$, $2$, or $3$. Furthermore,
$x^{\mu}$ will represent the set $\left( x^0, x^1, x^2, x^3 \right)$, $e_\nu$ the set $\left( e_0, e_1, e_2, e_3 \right)$, 
$g_{\mu \nu}$ a rank 2 covariant tensor, $g^{\mu \nu}$ a rank 2 contravariant tensor, and $g^{\mu}_{\nu}$ a rank 2 mixed tensor.
We will also use the Einstein summation convention for repeated indices; for example,
\begin{equation}
x^\mu g_{\mu \nu} = \sum\limits_{\mu = 0}^{3} x^\mu g_{\mu \nu}, \quad \nu = 0, 1, 2, 3, \quad \quad x^i  e_i = \sum\limits_{i=1}^3 x^i  e_i. \label{min3}
\end{equation} 

The Minkowski space is a foundation for building the mathematical structure of special relativity and the Minkowski metric,
\begin{equation}
\eta_{\mu \nu} = \begin{bmatrix} \ 1\ & \ 0\ & \ 0\ & \ 0\ \\ \ 0\ & \ -1\ & \ 0\ & \ 0\ \\
 \ 0\ & \ 0\ & \ -1\ & \ 0\ \\ \ 0\ & \ 0\ & \ 0\ & \ -1 \ \end{bmatrix} \label{min4}
\end{equation}
is an integral part of the formulation. $\eta_{\mu \nu}$ has the properties of a metric tensor discussed in Section \ref{sub:metric}.
It is a symmetric tensor and its inverse $\eta_{\mu \nu}^{-1}$ is $\eta^{\mu \nu}$ which is exactly the same as the right hand side
of \eqref{min4}. The metric tensor does not depend on the choice of a coordinate system.
The role of the metric tensor is to lower or raise the index of a vector, converting it from a contravariant vector to covariant
vector, or vice-versa. The metric tensor is also used for raising or lowering the indices of a tensor. For example,
\begin{equation}
x_\mu = \eta_{\mu \nu} x^\nu \quad \Longrightarrow \quad \braket{x_\mu \mid x^\mu} = c^2t^2 - x^2 - y^2 - z^2, \label{min5}
\end{equation}
for the Cartesian coordinate system. Note that the inner product can be negative; consequently, the Minkowski vector space
is considered to be a {\it pseudo inner product space}. Compared to regular vector spaces discussed earlier,
the Minkowski vector space does not have a unique zero (null) vector nor does it satisfy the triangle inequality.

For two infinitesimally close events,
\begin{equation}
x^\mu = \left( ct, x, y, z \right), \ {\rm  and} \  x^\mu + dx^\mu = \left( c \left(t + dt \right), x + dx, y + dy, z + dz \right), \label{min6}
\end{equation}
the square of the differential path length in Cartesian space is,
\begin{equation}
ds^2 =  c^2 dt^2 - dx^2 - dy^2 - dz^2 = dx^\mu \eta_{\mu \nu} dx^\nu. \label{min7}
\end{equation}
The two events are spacelike if $ds^2 < 0$, timelike if $ds^2 > 0$, and lightlike if $ds^2 = 0$. Timelike events and lightlike events can be
connected causally; this is not true for spacelike events. In special relativity, $ds^2$ is invariant with respect to different inertial frames; the
4-vectors corresponding to  the two events could be different for different inertial observers, but the space-time separation will be the same.
Consider two different inertial observers $O$ and $O'$ with 4-vectors $x^\mu$ and $x^{\prime \mu}$, respectively, representing
the same event. These two 4-vectors are connected by a linear transformation -- {\it Lorentz transformation} $\Lambda^\mu_\nu$,
\begin{equation}
x^{\prime \mu} = \Lambda^\mu_\nu x^\nu. \label{min8}
\end{equation}
The Lorentz transformation is a linear map representing a coordinate transform in the 4-vector Minkowski space,
\begin{equation}
\Lambda: {\mathscr M} \to {\mathscr M} \quad \Longrightarrow \quad x^\mu \mapsto x^{\prime \nu} = \Lambda^{\nu}_{\mu} x^\mu, \label{min8a}
\end{equation}
where $x^\mu, x^{\prime \nu} \in {\mathscr M}$.
From \eqref{min7} and \eqref{min8}, the invariance of $ds^2$ leads to,
\begin{equation}
\Lambda^{\mu}_\rho \eta_{\mu \nu} \Lambda^\nu_\xi = \eta_{\rho \xi}, \label{min9}
\end{equation}
or, equivalently, the matrix form,
\begin{equation}
\Lambda^{\rm T} \eta \Lambda = \eta. \label{min10}
\end{equation}
The determinant of \eqref{min10} gives $\left( {\rm det} \left( \Lambda \right) \right)^2 = 1$. Lorentz transformations with 
${\rm det} \left( \Lambda \right) = 1 $ are
referred to as {\it proper} while those with ${\rm det} \left( \Lambda \right) = -1$ as {\it improper} transformations. 
From \eqref{min9},
\begin{equation}
\eta_{0 0} = \Lambda^\mu_0 \eta_{\mu \nu} \Lambda^\nu_0 \quad \Longrightarrow \quad 
\left( \Lambda^0_0 \right)^2 = 1 + \sum\limits_{i=1}^3 \ \left( \Lambda^i_0 \right)^2 \ge 1, \label{min11}
\end{equation}
so that $\Lambda_0^0 \ge 1$ or $\Lambda_0^0 \le -1$. The former condition preserves the direction of time
and is referred to as {\it orthochronous}; the latter does not preserve the direction of time and is {\it non-orthochronous}.

Just for illustration, the Lorentz transformation (Lorentz boost) for an inertial frame $O'$ moving 
relative to $O$ with a uniform speed $v$ in the $x$-direction is,
\begin{equation}
\Lambda^{\mu}_\nu = \begin{bmatrix}
\ \gamma \ & \ - \gamma \beta \ & \ 0 \ & \ 0 \ \\
\ - \gamma \beta \ & \ \gamma \ & \ 0 \ & \ 0 \ \\
\ 0 \ & \ 0 \ & \ 1 \ & \ 0 \ \\
\ 0 \ & \ 0 \ & \ 0 \ & \ 1 \
\end{bmatrix}, \label{min12}
\end{equation}
where $\gamma = 1 / \sqrt{1 - \beta^2}$ and $\beta = v / c$.

Any scalar function $\phi$ of a 4-vector is invariant under coordinate transformations. Applying the chain rule for differentiation,
\begin{equation}
\frac{\partial \phi}{\partial x^{\prime \nu}} = \frac{\partial \phi}{\partial x^\mu} \, \frac{\partial x^\mu}{\partial x^{\prime \nu}}
\quad \Longrightarrow \quad  
\frac{\partial}{\partial x^{\prime \nu}} = \left( \Lambda^{-1} \right)^\mu_\nu \, \frac{\partial}{\partial x^\mu}.
\label{min12a}
\end{equation}
Following the discussion in Section \ref{subsub:contracov}, we note that partial derivatives transform like covariant vectors.
It is useful to write \eqref{min12a} in the form,
\begin{equation}
\partial_\nu = \left( \Lambda^{-1} \right)^\mu_\nu \partial^{\prime}_\mu, \quad {\rm where} \ \ \partial^{\prime}_\nu = \frac{\partial}{\partial x^{\prime \nu}},\ \ 
\partial_\mu = \frac{\partial}{\partial x^{\mu}}. \label{min12b}
\end{equation}

\subsection{\textbf{\textit{Covariant Form of Maxwell Equations}}}

The Maxwell equations in vacuum are,
\begin{align}
\nabla \cdot {\mathbf E}  & =  \frac{\rho_e}{\epsilon_0}, & 
\nabla \times {\mathbf B} -  \frac{1}{c^2} \frac{\partial}{\partial  t} {\mathbf E} & =
\mu_ 0 {\mathbf j}_e, \label{cfm0a} \\
\nabla \cdot {\mathbf B} & =  0, & 
\nabla \times {\mathbf E} + \frac{\partial}{\partial  t} {\mathbf B} & = 0. \label{cfm0b}
\end{align}
The right hand side of Gauss' law and the Ampere-Maxwell equation in \eqref{cfm0a} have the source terms and are referred to
as the inhomogeneous Maxwell equations. The Gauss' law for magnetism and the Faraday equation in \eqref{cfm0b} are the homogeneous
Maxwell equations. 

In the Cartesian Minkowski space, \eqref{cfm0a} can be written as,
\begin{equation}
\partial_\mu {F}^{\mu\nu} = \mu_0 J^\nu,\label{cfm1}
\end{equation}
where,
\begin{equation}
{F}^{\mu \nu} = \begin{bmatrix} \ 0\ & \ -E_x / c\ & \ -E_y / c\ & \ -E_z / c\ \\ 
E_x / c & 0 & -B_z & B_y \\ 
E_y / c & B_z & 0 & -B_x \\ 
E_z / c & -B_y & B_x & 0 
\end{bmatrix}, 
\label{cfm2}
\end{equation} 
is an anti-symmetric, rank 2, contravariant electromagnetic field tensor,
\begin{equation}
J^{\nu} = \left[ c \rho_e, \  j_{ex}, \  j_{ey}, \ j_{ez} \right], \label{cfm3}
\end{equation}
is a contravariant 4-vector  for charge and current densities.

The two homogeneous Maxwell equations \eqref{cfm0b} are expressed in terms
of elements of the covariant form of the electromagnetic field tensor $ {F}_{\alpha \beta}$,
\begin{equation}
\partial_\mu {F}_{\nu \rho} + \partial_\nu { F}_{\rho \mu} + \partial_\rho {F}_{\mu \nu} = 0, \label{cfm4}
\end{equation}
where,
\begin{equation}
{F}_{\alpha \beta}= \eta_{\alpha \mu} {F}^{\mu \nu} \eta_{\nu \beta} =
\begin{bmatrix} \ 0\ & \ E_x / c\ & \ E_y / c\ & \ E_z / c\ \\
-E_x / c & 0 & -B_z & B_y \\
-E_y / c & B_z & 0 & -B_x \\
-E_z / c & -B_y & B_x & 0 
\end{bmatrix}.
\label{cfm5}
\end{equation}
In  \eqref{cfm4}, $\mu$, $\nu$, and $\rho$ are distinctly different, and lead to four equations \eqref{cfm0b}. For example, 
if $\mu = 1$, $\nu = 2$, and $\rho = 3$, then \eqref{cfm4} leads to the Gauss' law for magnetism. The other values of the subscripts
$\mu$, $\nu$, and $\rho$ lead to the three components of the Faraday equation.

Although it takes a bit of algebra, it is possible to show that Maxwell equations in vacuum are Lorentz invariant.
However, at this stage, it is difficult to determine if the covariant form of Maxwell equations \eqref{cfm1} and \eqref{cfm4} can be put in a form
similar to the non-relativistic Schr\"odinger equation. The time derivative terms from \eqref{cfm1} and
\eqref{cfm4} lead to the state vector,
\begin{equation}
\ket{\psi} = \left[ E_x \ E_y \ E_z \ c B_x \ c B_y \ c B_z \right]^{\mathrm T}.
\end{equation}
Algebraic manipulations do not lead to the spatial variations having the form ${\mathbf H} \ket{\psi}$ as required by quantum postulate 3 in Section
\ref{pqm}. But, all is not lost. The following section describes an insightful choice of electromagnetic field variables for which the Maxwell equations
take on a form suitable for our purposes.

\section{Unitary Representation of Maxwell Equations in Vacuum}
\label{sec:urmev}

Let us define a new set of field variables,
\begin{equation}
{\mathbf F}^{\pm} \left( {\mathbf r}, t \right) \ = \ \frac{1}{\sqrt{2}} \; 
\left[ \sqrt{\epsilon_0} \; {\mathbf E} \left( {\mathbf r}, t \right)  
 \; \pm \; \frac{i}{\sqrt{\mu_0}} \; {\mathbf B} \left( {\mathbf r}, t \right) \right], \label{um1}
\end{equation} 
which are usually referred to as {\it Riemann-Silberstein-Weber} (RSW) vectors. It should be noted that ${\mathbf F}^+$
and ${\mathbf F}^-$ are not complex conjugate vectors since the electric and magnetic fields can be complex.
The four Maxwell equations take on the following form in terms of the RSW vectors,
\begin{align}
\nabla \cdot {\mathbf F}^{\pm} \left( {\mathbf r}, t \right) \ & = \ 
\frac{1}{\sqrt{2 \epsilon_0}} \; \rho_e, \label{um2} \\
i\; \frac{\partial {\mathbf F}^{\pm} \left( {\mathbf r}, t \right) }{\partial t} \ &= \ \pm \; c \;
\nabla \times {\mathbf F}^{\pm} \left( {\mathbf r}, t \right) 
- \frac{i}{\sqrt{2 \epsilon_0}} \; {\mathbf j}_e \label{um3}
\end{align}
If we choose the following state vector,
\begin{equation}
\ket{\psi \left( {\mathbf r}, t \right) } = \begin{bmatrix}
{-\mathrm F}_x^+ + i {\mathrm F}_y^+ \\
{\mathrm F}_z^+ \\
{\mathrm F}_z^+ \\
{\mathrm F}_x^+ + i {\mathrm F}_y^+ \end{bmatrix}, \label{um4}
\end{equation}
it follows from \eqref{um2} and \eqref{um3} that,
\begin{equation}
\gamma^\mu \partial_\mu  \ket{\psi \left( {\mathbf r}, t \right) } = {\mathbf J} \label{um5}
\end{equation}
where,
\begin{align}
\gamma^0 & = \begin{bmatrix} \ 1 \ & \ 0 \ & \ 0 \ & \ 0 \ \\
0 & 1 & 0 & 0 \\
0 & 0 & 1 & 0 \\
0 & 0 & 0 & 1 \end{bmatrix}, \quad \quad &
\gamma^1 & = \begin{bmatrix} \ 0 \ & \ 0 \ & \ 1 \ & \ 0 \ \\
0 & 0 & 0 & 1 \\
1 & 0 & 0 & 0 \\
0 & 1 & 0 & 0 \end{bmatrix}, \nonumber \\
\gamma^2 & = \begin{bmatrix} \ 0 \ & \ 0 \ & \ -i \ & \ 0 \ \\
0 & 0 & 0 & -i \\
i & 0 & 0 & 0 \\
0 & i & 0 & 0 \end{bmatrix}, \quad \quad &
\gamma^3 & = \begin{bmatrix} \ 1 \ & \ 0 \ & \ 0 \ & \ 0 \ \\
0 & 1 & 0 & 0 \\
0 & 0 & -1 & 0 \\
0 & 0 & 0 & -1 \end{bmatrix}, \label{um6} 
\end{align}
and,
\begin{equation}
{\mathbf J} = -\sqrt{\frac{\mu_0}{2}} \; \begin{bmatrix} -j_{ex} + i j_{ey} \\ j_{ez} - c \rho_e \\ j_{ez} + c \rho_e \\ j_{ex} + i j_{ey} \end{bmatrix},
\label{um6a}
\end{equation}
with ${\mathbf j}_e = \left( j_{ex}, j_{ey}, j_{ez} \right)$ in the Cartesian coordinate system.
The matrices $\gamma^\mu$ are all Hermitian and unitary, and have the following properties,
\begin{equation}
\left( \gamma^\mu \right) ^2 = {\mathcal I}, \nonumber 
\end{equation}
\begin{equation}
\gamma^1 \gamma^2 = - \gamma^2 \gamma^1 = i \gamma^3, 
\quad
\gamma^2 \gamma^3 = - \gamma^3 \gamma^2 = i \gamma^1, \quad
\gamma^3 \gamma^1 = - \gamma^1 \gamma^3 = i \gamma^2. \label{um7}
\end{equation}
If we ignore the external charge and current densities,
equation \eqref{um5} can be cast in the form of a Schr\"odinger equation,\footnote{The equation also
resembles the Dirac equation for a massless particle.}
\begin{equation}
\frac{1}{c} \; \frac{\partial}{\partial t} \ket{\psi} = - \; \gamma^i \; \frac{\partial}{\partial x^i} \; \ket{\psi}. \label{um8}
\end{equation}
This equation for wave propagation in vacuum is suitable for implementing on a quantum computer.

If we want to describe wave propagation in a simple uniform dielectric, the speed of light in vacuum $c$ is replaced by $v$ -- the
speed of light in the dielectric medium. For a medium with permittivity $ \epsilon $, $v = 1 / \sqrt{\epsilon \mu_0}$.

%% file: qla_oned.tex
\section{Quantum Lattice Algorithm for Maxwell Equations}
\label{sec:qla}

In this section, we will formulate a {\it quantum lattice algorithm} (QLA) which can be implemented on existing 
supercomputers and tested for accuracy and speed. The algorithm will be for propagation in the two-dimensional
$x-y$ plane. The configuration space is covered by a uniform lattice of discrete points. The electromagnetic fields
are prescribed at one edge of the spatial box at time $t = 0$, and the fields are evolved along the grid
as a function of time. We will assume that there are no external sources of charge and current densities.

The electromagnetic field vector is represented in terms of qubits,
\begin{equation}
\ket{\psi \left( x, y, t \right)} =
\begin{bmatrix} q_0 \left( x, y, t \right) \\
q_1 \left( x, y, t \right) \\ q_2 \left( x, y, t \right) \\ q_3 \left( x, y, t \right) \end{bmatrix}. \label{qla1}
\end{equation}
The relation of the qubits with the electromagnetic fields is obtained from \eqref{um4}. The space-time
evolution of the qubits is given by \eqref{um8} which, along with \eqref{um6}, gives,
\begin{equation}
\frac{\partial}{\partial t} \begin{bmatrix} \ q_0 \  \\ q_1 \\ q_2 \\ q_3 \end{bmatrix} = 
- \frac{\partial}{\partial x} \begin{bmatrix} \ q_2 \  \\ q_3 \\ q_0 \\ q_1 \end{bmatrix}
+ i \; \frac{\partial}{\partial y} \begin{bmatrix} \ q_2 \  \\ q_3 \\ - q_0 \\ -q_1 \end{bmatrix}. \label{qla2}
\end{equation}
In this equation we have replaced $v t$ by $t$, so that $x$, $y$, and $t$ have the same dimensions.
The evolution equation \eqref{um8} is linear and, consequently, separable in the Cartesian coordinate 
system. In developing a QLA we can treat the $x$ and $y$ directions separately.

There are two essential steps to a QLA; the first step streams the qubits from one lattice site to a neighboring site,
and the second step ``entangles'' two qubits at a given lattice site. Consequently, we have a series of streaming
and entanglement operators
which propagate information about the electromagnetic fields along the two-dimensional lattice. In order to satisfy the quantum postulates,
the streaming and entanglement operators have to be unitary.
From \eqref{qla2}, we note that there is coupling between the qubits in the time derivative and in the space derivatives: $ q_0 \leftrightarrow q_2 $
and $ q_1 \leftrightarrow q_3 $. The unitary matrices that entangle the qubits at each lattice site are,
\begin{equation}
C_x = \begin{bmatrix} \ \cos \theta \ & \ 0 \ & \ \sin \theta \ & \ 0 \ \\
0 & \cos \theta & 0 & \sin \theta \\
- \sin \theta & 0 & \cos \theta &0 \\
0 & - \sin \theta & 0 & \cos \theta 
\end{bmatrix}, 
C_y = \begin{bmatrix} \ \cos \theta \ & \ 0 \ & \ i \sin \theta \ & \ 0 \ \\
0 & \cos \theta & 0 & i \sin \theta \\
i \sin \theta & 0 & \cos \theta &0 \\
0 & i \sin \theta & 0 & \cos \theta 
\end{bmatrix}, \label{qla3}
\end{equation}
where $\theta$ is an entanglement angle that couples two qubits.

The streaming operation in itself is a two-step process; the first step streams qubits $q_0$ and $q_1$  while leaving $q_2$ and $q_3$ unchanged, and
the second step streams $q_2$ and $q_3$ leaving $q_0$ and $q_1$ unchanged. The two pertinent streaming operators in the $x$-direction 
are,
\begin{align}
S^{01}_{\pm x} 
\begin{bmatrix}
q_0 \left( x, y, t \right) \\ q_1 \left( x, y, t \right) \\ q_2 \left( x, y, t \right) \\ q_3 \left( x, y, t \right) \end{bmatrix} & =
\begin{bmatrix}
q_0 \left( x \pm dx, y, t \right) \\ q_1 \left( x \pm dx, y, t \right) \\ q_2 \left( x, y, t \right) \\ q_3 \left( x, y, t \right) \end{bmatrix}, \nonumber \\
S^{23}_{\pm x} 
\begin{bmatrix}
q_0 \left( x, y, t \right) \\ q_1 \left( x, y, t \right) \\ q_2 \left( x, y, t \right) \\ q_3 \left( x, y, t \right) \end{bmatrix} & =
\begin{bmatrix}
q_0 \left( x , y, t \right) \\ q_1 \left( x , y, t \right) \\ q_2 \left( x \pm dx, y, t \right) \\ q_3 \left( x \pm dx, y, t \right) \end{bmatrix}, \label{qla4}
\end{align}
where $\pm dx$ corresponds to a step of one lattice site in the $+x$ or $-x$ direction.
Similarly, we can set up the streaming operators in the $y$-direction: $S^{01}_{\pm y}$ and $S^{23}_{\pm y}$. 

The interleaving of the entanglement and streaming operators in the $x$-direction is constructed as follows,
\begin{align}  
U_x & = S^{01}_{-x} C_x S^{01}_{+x} C_x^\dagger\cdot S^{23}_{+x} C_x S^{23}_{-x} C_x^\dagger, \nonumber \\
\widetilde{U}_x & = S^{01}_{+x} C_x^\dagger S^{01}_{-x} C_x \cdot S^{23}_{-x} C_x^\dagger S^{23}_{+x} C_x. \label{qla5}
\end{align}  
Since the $y$-term in \eqref{qla2} has $i$ multiplying it, the $y$ sequence is a bit different,
\begin{align} 
U_y & = S^{23}_{-y} C_y S^{23}_{+y} C_y^\dagger \cdot S^{01}_{+y} C_y S^{01}_{-y} C_y^\dagger, \nonumber \\
\widetilde{U}_y & = S^{23}_{+y} C_y^\dagger S^{23}_{-y} C_y \cdot S^{01}_{-y} C_y^\dagger S^{01}_{+y} C_y.  \label{qla8}
\end{align} 

Finally, the time advancement from time $t$ to $t + \delta t$ ($\delta \ll 1$) is given by,
\begin{equation}
\begin{bmatrix} \ q_0 \  \\  q_1 \\  q_2 \\  q_3 \end{bmatrix}_{t + \delta t} =
\widetilde{U}_y U_y \widetilde{U}_x U_x 
\begin{bmatrix} \ q_0 \  \\ q_1 \\ q_2 \\ q_3 \end{bmatrix}_{t}. \label{qla10}
\end{equation}

\paragraph{\bf Continuum Limit of the Quantum Lattice Algorithm}

We need to verify that the discretization scheme used in the QLA is an appropriate representation of the
continuum equation; in other words, is \eqref{qla10} approximately the same as \eqref{um8}?

If $\varepsilon \ll 1$ is an ordering parameter, we approach the continuum limit by
setting the entanglement angle $\theta \sim \varepsilon$ and $dx, dy \sim \varepsilon$. Then,
\begin{equation} 
U_x  \approx {\mathcal I}_4 - \frac{\varepsilon^2}{2} \; \begin{bmatrix} \ 0_2\ & \ {\mathcal I}_2\ \\ {\mathcal I}_2 & 0_2 \end{bmatrix} \; 
\frac{\partial}{\partial x} \ + \ 
{\mathcal O} \left( \varepsilon ^3 \right), \label{qla6}
\end{equation} 
where ${\mathcal I}_n$ is the $n$-dimensional identity matrix and $0_2$ is a two-dimensional null matrix.  
We symmetrize the sequence to get a second-order accurate algorithm,  
\begin{equation}
{\widetilde U}_x U_x \approx {\mathcal I}_4 - \varepsilon^2 \; \begin{bmatrix} \ 0_2\ & \ {\mathcal I}_2\ \\ {\mathcal I}_2 & 0_2 \end{bmatrix} \; 
\frac{\partial}{\partial x} \ + \ 
{\mathcal O} \left( \varepsilon ^4 \right). \label{qla7}
\end{equation}  
Similarly,
\begin{equation}
{\widetilde U}_y U_y \approx {\mathcal I}_4 + i \varepsilon^2 \; \begin{bmatrix} \ 0_2\ & \ {\mathcal I}_2\ \\ - {\mathcal I}_2 & 0_2 \end{bmatrix} \; 
\frac{\partial}{\partial y} \ + \ 
{\mathcal O} \left( \varepsilon ^4 \right). \label{qla9}
\end{equation}
The small $\varepsilon$ expansions lead to,
\begin{equation}
\frac{\partial}{\partial t} \begin{bmatrix} \ q_0 \ \\ q_1 \\ q_2 \\ q_3 \end{bmatrix} =
- \frac{\partial}{\partial x} \begin{bmatrix} \ q_2 \ \\ q_3 \\ q_0 \\ q_1 \end{bmatrix}
+ \; i \frac{\partial}{\partial x} \begin{bmatrix} \ q_2 \ \\ q_3 \\ -q_0 \\ - q_1 \end{bmatrix} + {\mathcal O} \left( \varepsilon^2 \right),
\label{qla9a}
\end{equation}
which, to order $\varepsilon^2$ is the same as \eqref{qla2}.\footnote{$\varepsilon^2$ has been factored out to get \eqref{qla9a}.}

%% file: conclusions.tex
\section{Concluding Thoughts}
\label{sec:conc}

The Maxwell equations for electromagnetic waves in vacuum can be structured to be suitable for quantum
computers by an appropriate choice of the field variables. While this choice of variables is not necessarily unique, the
Riemann-Silberstein-Weber vectors are quite appropriate for waves in a uniform dielectric. 
Once we have Maxwell equations in the form of a Schr\"odinger equation, we
are able to construct a quantum lattice algorithm in terms of qubits. The streaming and entanglement operators
in the quantum lattice algorithm are unitary and reproduce Maxwell equations up to second order in the
lattice spacing. The extension of this modeling to a magnetized plasma is of keen interest.

The linear response of a magnetized plasma is dependent on the frequency of the electromagnetic wave. If the
electromagnetic fields are oscillating in time with an angular frequency $\omega$, the time dependence of all
the fields is of the form $\exp \left( - i \omega t \right)$. Suppose that the plasma is immersed in an
ambient magnetic field ${\mathbf B}_0 = B_0 \hat{\mathbf z}$ where $B_0$ is the magnitude of the magnetic
field and $\hat{\mathbf z}$ is a unit vector along the Cartesian $z$-direction. For a thermally cold plasma, 
the permittivity is a second rank tensor,
\begin{equation}
\epsilon_{ij} = \epsilon_0 \ 
\begin{pmatrix} 1 + \chi_{_{11}}& -i \chi_{_{12}} & 0 \\
i \chi_{_{12}}  & 1 + \chi_{_{11}} & 0 \\
0 & 0 & 1 + \chi_{_{33}} \end{pmatrix}
\label{cldpl1},
\end{equation}
where,
\begin{align}
\chi_{_{11}}\left( {\mathbf r},  \omega \right) & = 
  - \frac{\omega_{pe}^2 }{\omega^2 - \omega_{ce}^2} - \sum_i  \frac{\omega_{pi}^2 }{\omega^2 - \omega_{ci}^2}, \nonumber \\
\chi_{_{12}}\left( {\mathbf r},  \omega \right) & = 
 - \frac{ \omega_{ce}}{\omega} \frac{\omega_{pe}^2 }{\omega^2 - \omega_{ce}^2} + \sum_i  \frac{ \omega_{ci}}{\omega}
\frac{\omega_{pi}^2 }{\omega^2 - \omega_{ci}^2}, \label{cldpl3}\\
\chi_{_{33}}\left( {\mathbf r},  \omega \right) & =
 - \frac{\omega_{pe}^2}{\omega^2} - \sum_i  \frac{\omega_{pi}^2}{\omega^2}. \nonumber
\end{align}
In this expression the summation is over all the ion species in the plasma; the $i$-th ion has charge $Z_i |e|$ and mass $m_i$. 
The electron and ion angular plasma frequencies are $\omega_{pe} = \sqrt{e^2 n_e/\epsilon_0 m_e}$ and 
$\omega_{pi} = \sqrt{Z_i^2e^2 n_i/\epsilon_0 m_i}$, respectively; $n_i$ is the density of the $i$-th ion species.
The electron and ion angular cyclotron frequencies are $\omega_{ce} = |e| B_0 / m_e$ and $\omega_{ci} = Z_i |e| B_0 / m_i$, respectively.
The spatial dependence of the permittivity tensor is through the density.
The electric displacement field is $D_i = \epsilon_{ij} E_j$ with $\epsilon_{ij}$ being a Hermitian matrix.

If we could formulate an equivalent Schr\"odinger equation, it would be an eigenvalue equation since the time dependence of the fields is prescribed.
Consequently, we will not be able to study the temporal propagation of waves.

In a plasma, electromagnetic waves exchange energy with electron and ions. From the perspective of a wave, its amplitude decreases if it imparts
energy to the particles, while its amplitude grows if the energy flows from the particles to the wave. Concurrently, particles exchange energy through
elastic collisions. Since quantum mechanics deals with closed
systems in which total energy is conserved, developing a computational model with self-consistent
interaction of electromagnetic waves and charged particles in a plasma is a difficult task. 
Is it possible to simplify the self-consistent interaction by a constructing a closed system made up of two parts; one part which solves Maxwell equations
in a plasma while the other interacting part is either a sink or source of energy?

We dealt with linear wave propagation by assuming that the response of the plasma -- its permittivity -- is a linear function of the
electromagnetic field. This assumption allowed us to connect Maxwell equations with quantum mechanics as the Schr\"odinger equation
is a linear equation. However, intense electromagnetic fields can modify the permittivity of a plasma and induce a nonlinear dependence
on the intensity of the field. Such is the case, for example, in laser-plasma interactions. 
How can we incorporate nonlinear effects into a quantum representation of electromagnetic wave
propagation in nonlinear dielectrics?
Just to add to the challenge, the fluid description of a plasma, subject to electromagnetic fields, includes a convective derivative which is generally
nonlinear. A quantum representation of the fluid model will be useful for implementing on a quantum computer.

These are exciting times as we explore different and innovative strategies 
for molding classical plasma physics into a form suitable for quantum computations.
By being able to implement and test the algorithms resulting from these studies in present-day classical computers,
the wait for appropriate quantum computers is exhilarating. 

\paragraph{\bf Acknowledgement}

We are deeply grateful to Ms. Julianna Mullen for a careful reading of this manuscript and proposing changes that have enhanced the
clarity and flow of our narrative. We are also thankful to Professor Athanasios N. Yannacopoulos for his critical review of this chapter.
We are grateful to Dr. Didier Benisti for reading through the chapter and identifying areas of improvement.

A substantial fraction of our research on the application of quantum information science to 
classical wave propagation in plasmas is supported by the United States Department of Energy (Grant numbers
DE-SC0021647, DE-FG02-91ER-54109, and DE-SC0021651),
and by the EUROfusion Consortium, funded by the European Union via the Euratom Research and Training Programme 
(Grant Agreement No 101052200 -- EUROfusion).

%% file: appendix_A.tex
\section{Appendix}
\label{appendix}

\subsection{\textbf{\textit{Euclidean Space}}}
 
The fundamental objects in a Euclidean $n$-space, ${\mathbb E}^n$, are {\it points} which are $n$-tuples of real numbers 
$X = \left\{ x_1, x_2, \dots , x_n \mid x_i \in {\mathbb R}, i = 1, 2, \dots, n \right\}$. As an example, points in a one-dimensional
Euclidean space ${\mathbb E}^1$ lie on a line, while points in ${\mathbb E}^2$ lie in a plane.
As is the case for physical laws, a Euclidean space is devoid of any standard origin or any standard basis.
A Euclidean space is considered to be infinite and {\it flat} in which the axioms of Euclidean geometry are satisfied.  
Any two points in a Euclidean space are connected by a translation.
 
If $Y = \left\{ y_1, y_2, \dots , y_n \mid y_i \in {\mathbb R}, i = 1, 2, \dots, n \right\} \in {\mathbb E}^n$, then
\begin{itemize}
\item Equality: $X = Y$ if $x_i = y_i$ for all $i = 1, 2, \dots , n$
\item Addition: $X + Y = \left( x_1 + y_1, x_2 + y_2, \dots , x_n+ y_n \right)$ where $+$ is the usual arithmetic addition
\item Scalar multiplication: $a X = a \left( x_1, x_2, \dots , x_n \right) = \left( ax_1, ax_2, \dots , ax_n \right)$, where $a \in {\mathbb R}$
and $a x_i$ is the usual arithmetic product of two real numbers
\item Dot product: $ X \cdot Y = x_1 y_1 + x_2 y_2 + \dots + x_n y_n = \sum\limits_{i = 1}^n x_i y_i$
\item Distance function: $d \left(X , Y \right) = \sqrt{ \left( x_1 - y_1 \right)^2 + \left( x_2 - y_2 \right)^2 + \dots + \left( x_n - y_n \right)^2 }$.
\end{itemize}

\subsection{\textbf{\textit{Cartesian Space}}}

A {\it Cartesian space} is essentially a coordinate system within a Euclidean space. In Cartesian space, there exists an origin and we can set up 
a basis set so that any point in the Euclidean space can be expressed in terms of the basis set.  Thus, a Cartesian space in ${\mathbb R}^n$ is
a vector space with the origin as the zero vector. The inner product of any two vectors in ${\mathbb R}^n$, with an assigned basis set, is 
defined as in \eqref{nmtig} for real vectors.

\subsection{\textbf{\textit{Non-Euclidean Space, Tangent Space, and Local Description}}}

A non-Euclidean space is a space in which the rules of Euclidean space do not apply. We can essentially think of non-Euclidean spaces as spaces with
curvature. For example, along the surface of a sphere we cannot have two parallel lines or a triangle whose three angles add up to $180^\circ$ -- two
of the basic postulates of Euclidean space. We will assume that any non-Euclidean space is a {\it manifold} which is locally Euclidean.\footnote{We will
assume that the manifold is smooth and, hence, differentiable.} For example, a small
region around any given point on the surface of a three-dimensional sphere (the manifold) is locally flat -- the Euclidean space ${\mathbb E}^2$ being 
the tangent plane at that point. The generalization of a tangent line at a point on a curve, or a tangent plane at a point on a curved surface, to higher
dimensions is {\it tangent space}. For a $n$-dimensional manifold the tangent space is a $n$-dimensional vector space.

For a locally Euclidean space we can use the machinery of the flat Cartesian space. However, each point on a manifold will have different basis set.
In Cartesian space, the set of basis vectors is the same over all space. In Euclidean space, every coordinate system can be transformed into Cartesian
coordinate system. In non-Euclidean space, the basis vectors can be quite different (unequal magnitudes
and directions) along the manifold. For non-Euclidean spaces, it is convenient to consider differential elements within the tangent vector space just as in
calculus when we analyze continuous functions and their derivatives, or evaluate integrals.

Consider a vector ${\mathbf x} \left( \xi \right)$ that depends on a parameter $\xi$.\footnote{For example, $\xi$ could represent time.}
As $\xi$ changes, ${\mathbf  x}$ traces out a trajectory on
a manifold in a $n$-dimensions. At some point $P$ on the manifold,\footnote{Just as in Euclidean space, points are also
fundamental objects in non-Euclidean space.} any incremental changes in ${\mathbf x}$, due to an incremental change in $\xi$, 
is in the tangent space at $P$. The local tangent space being Euclidean, we assume that the coordinates of ${\mathbf x}$ depend
on $\xi$ but the Cartesian basis set is independent of $\xi$. Thus,
\begin{equation}
{\mathbf x} \left( \xi \right) = x^1\left( \xi \right) {\mathbf e}_1+ x^2\left( \xi \right) {\mathbf e}_2 + \dots + x^n\left( \xi \right) {\mathbf e}_n, \label{apenx1}
\end{equation}
where $\left\{ {\mathbf e}_1, {\mathbf e}_2, \dots , {\mathbf e}_n \right\}$ is a local Cartesian basis set.  
Then, for a small change $\delta \xi$ in $\xi$ within the tangent plane around $P$, the change in ${\mathbf x}$ is,
\begin{align}
\delta {\mathbf x} \left( \xi \right) & =   {\mathbf x} \left( \xi + \delta \xi \right) - {\mathbf x} \left( \xi \right) \nonumber \\
& \approx   \left( \frac{\delta x^1 \left( \xi \right)}{\delta \xi}  {\mathbf e}_1+ 
\frac{\delta x^2 \left( \xi \right)}{\delta \xi}  {\mathbf e}_2+  \dots + 
 \frac{\delta x^n \left( \xi \right)}{\delta \xi}  {\mathbf e}_n \right) \, \delta \xi, \label{apenx2}
\end{align}
where $ \delta x^i(\xi)/ \delta \xi$ is the derivative of $x^i$ with respect to $\xi$.
For a different set of Cartesian basis vectors, $\left\{ \tilde{\mathbf e}_1, \tilde{\mathbf e}_2, \dots , \tilde{\mathbf e}_n \right\}$,
\begin{equation}
{\mathbf x} \left( \xi \right) = \tilde{x}^1\left( \xi \right) \tilde{\mathbf e}_1+ \tilde{x}^2\left( \xi \right) \tilde{\mathbf e}_2 + \dots + 
\tilde{x}^n\left( \xi \right) \tilde{\mathbf e}_n, \label{apenx3}
\end{equation}
Following the notation in Section \ref{sub:transform}, the two basis sets are connected by a transformation tensor,\footnote{We follow the
Einstein summation convention.}
\begin{align}
\tilde{\mathbf e}_j  & = {\mathrm R}^i_j \, {\mathbf e}_i,   & {\mathbf e}_i  & = \left( {\mathrm R}^{-1} \right)^j_i \, \tilde{\mathbf e}_j, \label{apenx4}\\
\tilde{x}^j  & = \left( {\mathrm R}^{-1} \right)^j_i \, {x}^i, &  {x}^i & = {\mathrm R}^i_j \, \tilde{x}^j, \label{apenx5}
\end{align}
where $i, j = 1, 2, \dots, n$.
The partial derivatives of \eqref{apenx5} yield,
\begin{equation}
\frac{\partial \tilde{x}^j}{\partial {x}^i}=\left( {\mathrm R}^{-1} \right)^j_i, \quad\quad \frac{\partial x^k}{\partial \tilde{x}^l}= {\mathrm R}^k_l. \label{apenx6}
\end{equation}
From the chain rule for partial derivatives,
\begin{equation}
\frac{\partial \tilde{x}^j}{\partial {x}^i} \, \frac{\partial x^i}{\partial \tilde{x}^l} =\delta^j_l. \label{apenx7}
\end{equation}
If we substitute the right hand side of the two equations \eqref{apenx6} in the left hand side of \eqref{apenx7}, we obtain the same identity,
\begin{equation}
\left( {\mathrm R}^{-1} \right)^j_i \,  {\mathrm R}^i_l = \delta^j_l. \label{apenx8}
\end{equation}
Thus, we can obtain the transformation operator from the partial derivates of the coordinates in the two basis sets.

Consider a contravariant vector ${\mathbf v} \left( X \right)$ in the basis set $\left\{ {\mathbf e}_i \right\}$ $\left( i = 1, 2, \dots , n  \right) $, 
with coordinates $\left\{ v^1\left( X \right), v^2\left( X \right), \dots, v^n\left( X \right) \right\}$ 
that are functions of $X = \left\{ x^1, x^2, \dots , x^n \right\}$. The corresponding coordinates of ${\mathbf v}$ in the
basis set $ \left\{ \tilde{\mathbf e}_i \right\}$ are given by,
\begin{equation}
\tilde{v}^j \left( \widetilde{X} \right) = \frac{\partial \tilde{x}^j}{\partial x^i}\ v^i \left( X \right), \label{apenx9}
\end{equation}
where $\widetilde{X} = \left\{ \tilde{x}^1, \tilde{x}^2, \dots , \tilde{x}^n \right\}$. Equation \eqref{apenx9} gives the transformation rule
for contravariant vectors. The transformation rule for contravariant vectors can be extended to a contravariant tensor of rank $s$,
\begin{equation}
\widetilde{T}^{j_1, j_2, \dots , j_s} \left( \widetilde{X} \right) = \frac{\partial \tilde{x}^{j_1}}{\partial x^{i_1}} \, 
\frac{\partial \tilde{x}^{j_2}}{\partial x^{i_2}} \, \dots \, \frac{\partial \tilde{x}^{j_s}}{\partial x^{i_s}} \ T^{i_1, i_2, \dots , i_s}, \label{apenx9a}
\end{equation}
where $s \le n$ is an integer and $i_s , j_s = 1, 2, \dots, n$.

Covariant vectors are defined in dual vector space -- dual to the vector space of contravariant vectors. Here we will follow the discussion in
Sections \ref{subsub:contracov} and \ref{sub:transform}. For a covariant vector,
\begin{equation}
{\mathfrak g} \left( X \right) = g_1 \left( X \right) {\mathbf f}^1 + g_2 \left( X \right) {\mathbf f}^2 + \dots + g_n \left( X \right) {\mathbf f}^n,
\label{apenx10}
\end{equation}
the transformation rule is,
\begin{equation}
\tilde{g}_j \left( \widetilde{X} \right) = \frac{\partial x^i}{\partial \tilde{x}^j} \ g_i \left( X \right). \label{apenx11} 
\end{equation}
From the chain rule \eqref{apenx7}, we notice that the transformation for the covariant vector \eqref{apenx11} is the inverse of the 
transformation for the contravariant vector \eqref{apenx9}. This relationship is the same as in Section \ref{sub:transform}.
The transformation of covariant and mixed tensors can be expressed in a form similar to that of \eqref{apenx9a}.

%% file: supplementary_mat.tex
\section{Bibliography}
\label{sec:supp}

For different topics addressed in this chapter we include a list of references. This is an abbreviated list as there
are numerous books on each subject indicated below.

\paragraph{\bf Plasma Physics}

The material on plasma physics in Sections \ref{sec:intro} and \ref{sec:conc} is obtained
\begin{itemize}
\item Abraham Bers, {\it Plasma Physics and Fusion Plasma Electrodynamics,} Volume 1, Oxford University Press (2016).
\end{itemize}

\paragraph{\bf Linear Algebra}

There are many superb books on linear algebra and functional analysis. Among them are the following,

\begin{itemize}
\item Sheldon Axler, {\it Linear Algebra Done Right }, Fourth Edition, Springer, New York (2024).

\item Ray M. Bowen and Chow-Cheng Wang, {\it Introduction to Vectors and Tensors}, Volumes 1 \& 2, Dover Publications,
New York (2009).

\item P. K. Jain, O. P. Ahuja, and Khalil Ahmad, {\it Functional Analysis}, John Wiley, New York (1995).

\item Nadir Jeevanjee, {\it An Introduction to Tensors and Group Theory for Physicists,} Second Edition,
Birkh\"auser, Boston (Springer International, New York) (2015).

\item Serge Lang, {\it Linear Algebra}, Addison-Wesley, Reading (Massachusetts), (1968).

\item Michael Reed and Barry Simon, {\it Methods of Modern Mathematical Physics, Functional Analysis}, Academic Press, 
San Diego (1980). 

\item  Steven Roman, {\it Advanced Linear Algebra}, Second Edition, Springer-Verlag, New York (2008).

\end{itemize}

\paragraph{\bf Quantum Mechanics}

For further reading on quantum postulates, Hilbert spaces, Dirac notation, and quantum mechanics,

\begin{itemize}

\item Michel Le Bellac, {\it Quantum Physics}, Cambridge University Press, Cambridge (2006).

\item Robert B. Griffiths, {\it Consistent Quantum Theory}, Cambridge University Press, Cambridge (2002).

\item Chris J. Isham, {\it Lectures on Quantum Theory: Mathematical and Structural Foundations}, Imperial College Press, London (1995). 

\end{itemize}

\paragraph{\bf Quantum Computing}

\begin{itemize}

\item Michael A. Nielsen and Isaac L. Chuang, {\it Quantum Computation and Quantum Information}, $10^{th}$ Anniversary Edition, Cambridge
University Press, Cambridge (2010).

\item Eleanor Rieffel and Wolfgang Polak, {\it Quantum Computing: A Gentle Introduction}, Massachusetts Institute of Technology Press, Cambridge (2011).

\end{itemize}

\paragraph{\bf Electrodynamics and Relativity}

\begin{itemize}

\item Charles A. Brau, {\it Modern Problems in Classical Electrodynamics}, Oxford University Press, Oxford (2004).

\item David J. Griffiths, {\it Introduction to Electrodynamics}, Fifth Edition, Cambridge University Press, Cambridge (2023).

\item John D. Jackson, {\it Classical Electrodynamics}, Third Edition, Wiley, New Jersey (1998).

\item Andrew Zangwill, {\it Modern Electrodynamics}, Cambridge University Press, Cambridge (2012).

\end{itemize}

\paragraph{\bf Unitary Representation of Maxwell Equations}

The material in Section \ref{sec:urmev} is based on,
\begin{itemize}

\item Sameen Ahmed Khan, ``An exact matrix representation of Maxwell's equations,''
{\it Physica Scripta} {\bf 71}, 440-442 (2005).

\end{itemize}
 
\noindent The following articles are on the Riemann-Silberstein-Weber vectors, 

\begin{itemize}

\item Iwo Bialynicki-Birula and Zofia Bialynicka-Birula, ``The role of the Riemann-Silberstein vector in classical and quantum theories of electromagnetism,'' 
{\it Journal of Physics A: Mathematical and Theoretical} {\bf 46}, 053001-1-053001-32 (2013). 

\item Michael K.-H. Kiessling and A. Shadi Tahvildar-Zadeh, ``On the quantum-mechanics of a single photon,'' {\it Journal of Mathematical Physics}
{\bf 59}, 112302-1-112303-33 (2018). 

\item Charles T. Sebens, ``Electromagnetism as Quantum Physics,'' {\it Foundations of Physics} {\bf 49}, 365-389 (2019). 

\end{itemize}

\paragraph{\bf Quantum Lattice Algorithm}

The quantum lattice algorithm is discussed in the following papers and various references therein,
\begin{itemize}
\item George Vahala, Linda Vahala, Min Soe,  and Abhay K. Ram, ``Unitary quantum lattice simulations for Maxwell equations in vacuum and in dielectric media,''
{\it Journal of Plasma Physics} {\bf 86} 905860518-1-905860518-17 (2020).

\item Linda Vahala, George Vahala, Min Soe, Abhay Ram, and Jeffrey Yepez,
``Unitary qubit lattice algorithm for three-dimensional vortex solitons in hyperbolic self-defocusing media,''
{\it Communications in Nonlinear Science and Numerical Simulation} {\bf  75}, 152-159 (2019).

\end{itemize}

\paragraph{\bf Recent Publications}

The following publications are on our research related to electromagnetic wave propagation in plasma,

\begin{itemize}

\item Efstratios Koukoutsis, Kyriakos Hizanidis, Abhay K. Ram, and George Vahala, 
``Quantum simulation of dissipation for Maxwell equations in dispersive media,'' 
{\it Future Generation Computer Systems} {\bf 159}, 221-229 (2024). 

\item Efstratios Koukoutsis, Kyriakos Hizanidis, George Vahala, Min Soe, Linda Vahala, and Abhay K. Ram,
``Quantum computing perspective for electromagnetic wave propagation in cold magnetized plasmas,''
{\it Physics of Plasmas} {\bf 30}, 122108-1-122108-11 (2023).

\item Efstratios Koukoutsis, Kyriakos Hizanidis, Abhay K. Ram, and George Vahala,
``Dyson maps and unitary evolution for Maxwell equations in tensor dielectric media,''
{\it Physical Review A} {\bf 107}, 042215-1-042215-10 (2023).

\item George Vahala, Min Soe, Linda Vahala, Abhay K. Ram, Efstratios Koukoutsis, and Kyriakos Hizanidis,
``Qubit lattice algorithm simulations of Maxwell’s equations for scattering from anisotropic dielectric objects,''
{\it Computers \& Fluids} {\bf 266} 106039-1-106039-9 (2023).

\item George Vahala, Min Soe, Linda Vahala, and Abhay K. Ram, 
``Two dimensional electromagnetic scattering from dielectric objects using quantum lattice algorithm,'' 
available at SSRN: https://ssrn.com/abstract=3996913 or \\
http://dx.doi.org/10.2139/ssrn.3996913 (2021). 

\end{itemize}